\documentclass[11pt,english]{revtex4}
\usepackage{lmodern}

\usepackage[T1]{fontenc}
\usepackage[latin9]{inputenc}
\usepackage{babel}

\usepackage{amsmath}
\usepackage{amssymb}
\usepackage{esint}
\usepackage[unicode=true, pdfusetitle,
 bookmarks=true,bookmarksnumbered=false,bookmarksopen=false,
 breaklinks=false,pdfborder={0 0 1},backref=false,colorlinks=false]
 {hyperref}

\makeatletter
\@ifundefined{textcolor}{}
{%
 \definecolor{BLACK}{gray}{0}
 \definecolor{WHITE}{gray}{1}
 \definecolor{RED}{rgb}{1,0,0}
 \definecolor{GREEN}{rgb}{0,1,0}
 \definecolor{BLUE}{rgb}{0,0,1}
 \definecolor{CYAN}{cmyk}{1,0,0,0}
 \definecolor{MAGENTA}{cmyk}{0,1,0,0}
 \definecolor{YELLOW}{cmyk}{0,0,1,0}
 }

\usepackage{latexsym}\usepackage{bm}

\makeatother

\makeatother

\begin{document}

\title{Unitarity analysis of general Born-Infeld gravity theories}

\author{\.{I}brahim Güllü }

\email{e075555@metu.edu.tr}

\affiliation{Department of Physics,\\
 Middle East Technical University, 06531, Ankara, Turkey}

\author{Tahsin Ça\u{g}r\i{} \c{S}i\c{s}man}

\email{sisman@metu.edu.tr}

\affiliation{Department of Physics,\\
 Middle East Technical University, 06531, Ankara, Turkey}

\author{Bayram Tekin}

\email{btekin@metu.edu.tr}

\affiliation{Department of Physics,\\
 Middle East Technical University, 06531, Ankara, Turkey}

\date{\today}
\begin{abstract}
We develop techniques of analyzing the unitarity of general Born-Infeld
gravity actions in D-dimensional spacetimes. The determinantal form
of the action allows us to find a compact expression quadratic in
the metric fluctuations around constant curvature backgrounds. This
is highly nontrivial since for the Born-Infeld actions, in principle,
infinitely many terms in the curvature expansion should contribute
to the quadratic action in the metric fluctuations around constant
curvature backgrounds, which would render the unitarity analysis intractable.
Moreover in even dimensions, unitarity of the theory depends only
on finite number of terms built from the powers of the curvature tensor.
We apply our techniques to some four-dimensional examples. 

\tableofcontents{} 
\end{abstract}
\maketitle

\section{Introduction}

Tree-level unitarity analysis, that is tachyon and ghost freedom,
of a generic gravity model with arbitrary powers of the curvature
tensors around a constant (nonzero) curvature background is a nontrivial
problem. On the other hand, for flat backgrounds, only the quadratic
terms contribute to the propagators, and therefore the analysis is
rather simple. In fact, in four dimensions the only unitary model,
apart from the Einstein's gravity, is the $R+\alpha R^{2}$ theory
at the quadratic order. But, this model is not renormalizable without
a $\beta R_{\mu\nu}^{2}$ term which, when augmented to the action,
ruins unitarity by introducing a massive ghost mode \cite{Stelle}. 

Experience from quantum field theory dictates that at high energies
Einstein's gravity should be replaced with a theory that has higher
powers of various curvature tensors symbolically written in the form\begin{equation}
I=\int d^{4}x\,\left\{ \frac{1}{\kappa}\left(R-2\Lambda_{0}\right)+\sum_{n=2}^{\infty}a_{n}\left(\text{Riem, Ric, R, }\nabla\text{Riem, }\dots\right)^{n}\right\} .\label{eq:generic_higher_ord_act}\end{equation}
 The main nontrivial question is how to find the correct couplings
$a_{n}$ that yield a viable unitary theory. One might view gravity
as a low energy approximation to a microscopic theory such as string
theory and thus expect to find a unitary (but not necessarily renormalizable)
gravity theory to any desired order in the curvature by perturbatively
computing $a_{n}$. Of course beyond quadratic order this is a very
difficult computational problem. Another approach is the so called
asymptotically safe gravity which conjectures that the dimensionless
versions of all the coupling constants in (\ref{eq:generic_higher_ord_act})
have a nontrivial UV fixed point and even for infinitely many coupling
constants the theory has predictive power since the critical surface
is finite dimensional \cite{Weinberg1,Niedermaier,Weinberg2}. In
this work, encouraged by our recent observation in three dimensions
that we briefly summarize below, we take a different route and propose
that certain Born-Infeld (BI) type gravity actions might define unitary
models to all orders. Unitarity analysis around constant curvature
backgrounds is itself a complicated problem when many powers of curvature
tensors are involved, here we develop the techniques of carrying out
this analysis in detail and provide two nonunitary examples in four
dimensions. In subsequent work \cite{Gullu0}, we will give more examples
in three and four dimensions that are unitary.

Let us recapitulate the three-dimensional BI gravity action \cite{Gullu}
and its success of reproducing the known viable theories: \begin{equation}
I_{\text{BINMG}}=-\frac{4m^{2}}{\kappa^{2}}\int d^{3}x\,\left[\sqrt{-\det\left(g-\frac{1}{m^{2}}G\right)}-\left(1-\frac{\lambda}{2}\right)\sqrt{-\det g}\right],\label{eq:BI-NMG_action}\end{equation}
 where the components of the matrix $G$ read as $G_{\mu\nu}\equiv R_{\mu\nu}-\frac{1}{2}g_{\mu\nu}R$.
In (\ref{eq:BI-NMG_action}), we have referred to this model as the
BI extended new massive gravity (BINMG), since in small curvature
expansion, this model reproduces the cosmological Einstein-Hilbert
theory at the first order, the new massive gravity theory \cite{BHT},
which is unitary \cite{Deser,Nakasone,Gullu1,Gullu2}, at the second
order and the extended new massive gravity based on the existence
of the holographic $c$-functions at the cubic and fourth orders \cite{Sinha,Gullu}.
With the help of the techniques we develop below, we have shown that
BINMG is a unitary theory at all orders around flat and constant curvature
vacua \cite{Gullu0}. One of course would like to find analogs of
(\ref{eq:BI-NMG_action}) in higher, especially in four, dimensions.
To be able to do this, one has to first establish tools for the unitarity
analysis which is the purpose of this work. In what follows, for the
sake of generality, we will keep the discussion in $D$ dimensions
and for generic BI actions with the only restriction that they reproduce
the (cosmological) Einstein-Hilbert theory at the first order. 

The history of the BI-type actions is quite rich and for the nongravitational
cases a nice review was given in \cite{gibbons}. As for gravity,
BI-type gravitational actions actually precedes a decade their counterparts
in electrodynamics. It was Eddington who first proposed that, at least
in the absence of matter, using the connection as the independent
variable, Einstein-Hilbert action can be replaced by $I=\int d^{4}x\sqrt{\det R_{\mu\nu}\left(\Gamma\right)}$
\cite{Eddington}. (Note that one actually has to dig this result
out from Eddington's book, since it is not clearly stated in one place.
But, Schrödinger, attributing to Eddington, writes this action explicitly
on page 113 of his book \cite{Schrodinger}). More recently, Eddington's
approach (in fact a slight modification of it) was resuscitated in
\cite{BanadosFerreira} (and the references therein) as an alternative
to Big Bang cosmology without an initial singularity and with finite
density. In \cite{gibbonsDeser}, instead of Eddington's Palatini
formulation, the metric formulation where the metric is the only independent
variable was used in the form $I=\int d^{4}x\sqrt{\det\left(g_{\mu\nu}+\alpha R_{\mu\nu}+X_{\mu\nu}\right)}$
and constraints such as ghost freedom on BI-type gravity actions was
studied. Our work follows this line of thought and extends the unitarity
analysis to constant curvature spaces. We would like to point out
to some related works where BI-type gravities, their cosmological
and other solutions have been studied \cite{Wohlfarth,Nieto,ComelliDolgov,Comelli,Fiorini}.

The main idea of this work is to find a way to obtain the quadratic
action in the \emph{metric perturbation} of a generic BI gravity around
its constant curvature vacuum, and this can be achieved either by
explicitly calculating the $O\left(h_{\mu\nu}^{2}\right)$ action
or by finding the equivalent quadratic action in the \emph{curvature}
that has the same propagator with the original action. Once the equivalent
quadratic theory is known unitarity analysis follows with the conventional
methods described in \cite{Gullu1}. To facilitate understanding and
show what is to be expected, let us give one of our results here.
Let $A_{\mu\nu}$ be an \emph{arbitrary} $\left(0,2\right)$ tensor
built from the curvature tensors, then we will show that, in four
dimensions, at $O\left(h_{\mu\nu}\right)$ and $O\left(h_{\mu\nu}^{2}\right)$,
the action \begin{equation}
I=\frac{2}{\kappa\alpha}\int d^{4}x\,\left[\sqrt{-\det\left(g_{\mu\nu}+A_{\mu\nu}\right)}-\left(\alpha\Lambda_{0}+1\right)\sqrt{-\det g}\right],\end{equation}
 is equivalent to the simpler action \begin{align}
I_{O\left(A^{2}\right)} & =\frac{1}{\kappa\alpha}\int d^{4}x\,\sqrt{-g}\left(A-2\alpha\Lambda_{0}+\frac{1}{4}A^{2}-\frac{1}{2}A_{\mu\nu}^{2}\right),\end{align}
 where $A$ is the trace of $A_{\mu\nu}$. Once this is done unitarity
analysis can be carried out with the known methods which we shall
not repeat in this work.

The layout of the paper is as follows: In Sec. II, second order expansions
of the relevant tensors in the metric perturbation $h_{\mu\nu}$ are
given. Section III is the bulk of the paper which contains our general
analysis of BI gravities and the corresponding equivalent quadratic
actions. We also give two examples in four dimensions in this section.
Some technical details are delegated to the Appendices.

\section{Second Order Expansions of Curvature Tensors}

In order to study the fluctuations of generic BI actions around constant
curvature backgrounds, we will need to expand various tensors up to
second order in the metric perturbation $h_{\mu\nu}$ which is defined
as \begin{equation}
g_{\mu\nu}\equiv\bar{g}_{\mu\nu}+\tau h_{\mu\nu},\label{eq:Perturbation_def}\end{equation}
 where we introduced a small (dimensionless) parameter $\tau$ and
a background metric $\bar{g}_{\mu\nu}$ which is quite generic at
this stage (i.e. not necessarily constant curvature). {[}Taking the
risk of being pedantic, let us note that (\ref{eq:Perturbation_def})
is exact, and that there of course does not exist a natural dimensionless
parameter in gravity at all scales. So, what one actually means by
(\ref{eq:Perturbation_def}) is that in some frame $h_{\mu\nu}$ is
small compared to $\bar{g}_{\mu\nu}$ for all points in the spacetime,
and since there will be another expansion, that is the curvature expansion,
$\tau$ is introduced to keep track of the $h_{\mu\nu}$ orders.{]}
Some of the computations in this section are actually somewhat tedious
but straightforward. They could also be found in the literature, albeit
somewhat scattered, and probably not in the form we present here which
proved quite handy in our calculations that follow in the remainder
of this work. The inverse metric $g^{\mu\nu}$ can be found as\begin{equation}
g^{\mu\nu}=\bar{g}^{\mu\nu}-\tau h^{\mu\nu}+\tau^{2}h^{\mu\rho}h_{\rho}^{\nu}+O\left(\tau^{3}\right).\label{eq:Oh2_of_inv_met}\end{equation}
 The trace of the metric perturbation is given as $h=\bar{g}^{\mu\nu}h_{\mu\nu}$.
By using these results, the second order expansion of the Christoffel
connection becomes\begin{equation}
\Gamma_{\mu\nu}^{\rho}=\bar{\Gamma}_{\mu\nu}^{\rho}+\tau\left(\Gamma_{\mu\nu}^{\rho}\right)_{L}-\tau^{2}h_{\beta}^{\rho}\left(\Gamma_{\mu\nu}^{\beta}\right)_{L}+O\left(\tau^{3}\right),\end{equation}
 where $\bar{\Gamma}_{\mu\nu}^{\rho}$ is a background metric compatible
connection $\bar{\nabla}_{\rho}\bar{g}_{\mu\nu}=0$ and the linearized
connection $\left(\Gamma_{\mu\nu}^{\rho}\right)_{L}$ is defined as\begin{equation}
\left(\Gamma_{\mu\nu}^{\rho}\right)_{L}\equiv\frac{1}{2}\bar{g}^{\rho\lambda}\left(\bar{\nabla}_{\mu}h_{\nu\lambda}+\bar{\nabla}_{\nu}h_{\mu\lambda}-\bar{\nabla}_{\lambda}h_{\mu\nu}\right).\label{eq:Linear_Christoffel}\end{equation}
 The main object to consider is the Riemann tensor from which all
the other curvature tensors and scalars follow. Hence, substitution
of $\Gamma_{\mu\nu}^{\rho}=\bar{\Gamma}_{\mu\nu}^{\rho}+\delta\Gamma_{\mu\nu}^{\rho}$
to the Riemann tensor $R_{\phantom{\mu}\nu\rho\sigma}^{\mu}\equiv\partial_{\rho}\Gamma_{\sigma\nu}^{\mu}+\Gamma_{\rho\lambda}^{\mu}\Gamma_{\sigma\nu}^{\lambda}-\rho\leftrightarrow\sigma$
yields \begin{equation}
R_{\phantom{\mu}\nu\rho\sigma}^{\mu}=\bar{R}_{\phantom{\mu}\nu\rho\sigma}^{\mu}+\bar{\nabla}_{\rho}\left(\delta\Gamma_{\sigma\nu}^{\mu}\right)-\bar{\nabla}_{\sigma}\left(\delta\Gamma_{\rho\nu}^{\mu}\right)+\delta\Gamma_{\rho\lambda}^{\mu}\delta\Gamma_{\sigma\nu}^{\lambda}-\delta\Gamma_{\sigma\lambda}^{\mu}\delta\Gamma_{\rho\nu}^{\lambda},\end{equation}
 where $\delta\Gamma_{\mu\nu}^{\rho}=\tau\left(\Gamma_{\mu\nu}^{\rho}\right)_{L}-\tau^{2}h_{\beta}^{\rho}\left(\Gamma_{\mu\nu}^{\beta}\right)_{L}$
at this order. Therefore, the Riemann tensor becomes\begin{align}
R_{\phantom{\mu}\nu\rho\sigma}^{\mu}= & \bar{R}_{\phantom{\mu}\nu\rho\sigma}^{\mu}+\tau\left(R_{\phantom{\mu}\nu\rho\sigma}^{\mu}\right)_{L}-\tau^{2}h_{\beta}^{\mu}\left(R_{\phantom{\mu}\nu\rho\sigma}^{\beta}\right)_{L}\nonumber \\
 & -\tau^{2}\bar{g}^{\mu\alpha}\bar{g}_{\beta\gamma}\left[\left(\Gamma_{\rho\alpha}^{\gamma}\right)_{L}\left(\Gamma_{\sigma\nu}^{\beta}\right)_{L}-\left(\Gamma_{\sigma\alpha}^{\gamma}\right)_{L}\left(\Gamma_{\rho\nu}^{\beta}\right)_{L}\right]+O\left(\tau^{3}\right).\label{eq:Second_order_Riemann}\end{align}
 Note that raising and lowering is done by $\bar{g}_{\mu\nu}$, but
in the above expression, for the sake of notational clarity, we do
not raise and lower the indices of the linearized Christoffel connection.
Here, the linearized Riemann tensor $\left(R_{\phantom{\mu}\nu\rho\sigma}^{\mu}\right)_{L}$
is defined as\begin{equation}
\left(R_{\phantom{\mu}\nu\rho\sigma}^{\mu}\right)_{L}\equiv\frac{1}{2}\left(\bar{\nabla}_{\rho}\bar{\nabla}_{\sigma}h_{\nu}^{\mu}+\bar{\nabla}_{\rho}\bar{\nabla}_{\nu}h_{\sigma}^{\mu}-\bar{\nabla}_{\rho}\bar{\nabla}^{\mu}h_{\sigma\nu}-\bar{\nabla}_{\sigma}\bar{\nabla}_{\rho}h_{\nu}^{\mu}-\bar{\nabla}_{\sigma}\bar{\nabla}_{\nu}h_{\rho}^{\mu}+\bar{\nabla}_{\sigma}\bar{\nabla}^{\mu}h_{\rho\nu}\right).\label{eq:Linear_Riemann}\end{equation}
 With this result, the second order expansion of the Ricci tensor
and the scalar curvature, respectively, take the following forms \begin{align}
R_{\nu\sigma}= & \bar{R}_{\nu\sigma}+\tau\left(R_{\nu\sigma}\right)_{L}-\tau^{2}h_{\beta}^{\mu}\left(R_{\phantom{\mu}\nu\mu\sigma}^{\beta}\right)_{L}\nonumber \\
 & -\tau^{2}\bar{g}^{\mu\alpha}\bar{g}_{\beta\gamma}\left[\left(\Gamma_{\mu\alpha}^{\gamma}\right)_{L}\left(\Gamma_{\sigma\nu}^{\beta}\right)_{L}-\left(\Gamma_{\sigma\alpha}^{\gamma}\right)_{L}\left(\Gamma_{\mu\nu}^{\beta}\right)_{L}\right]+O\left(\tau^{3}\right),\label{eq:Second_order_Ricci}\end{align}
 \begin{align}
R= & \bar{R}+\tau R_{L}+\tau^{2}\left\{ \bar{R}^{\rho\lambda}h_{\alpha\rho}h_{\lambda}^{\alpha}-h^{\nu\sigma}\left(R_{\nu\sigma}\right)_{L}-\bar{g}^{\nu\sigma}h_{\beta}^{\mu}\left(R_{\phantom{\mu}\nu\mu\sigma}^{\beta}\right)_{L}\right.\nonumber \\
 & \phantom{\bar{R}+\tau^{2}R_{L}+\tau^{2}}\left.-\bar{g}^{\nu\sigma}\bar{g}^{\mu\alpha}\bar{g}_{\beta\gamma}\left[\left(\Gamma_{\mu\alpha}^{\gamma}\right)_{L}\left(\Gamma_{\sigma\nu}^{\beta}\right)_{L}-\left(\Gamma_{\sigma\alpha}^{\gamma}\right)_{L}\left(\Gamma_{\mu\nu}^{\beta}\right)_{L}\right]\right\} +O\left(\tau^{3}\right),\label{eq:Second_order_R}\end{align}
 where the linearized Ricci tensor and the linearized scalar curvature
are defined, respectively, as \begin{equation}
R_{\nu\sigma}^{L}\equiv\frac{1}{2}\left(\bar{\nabla}_{\mu}\bar{\nabla}_{\sigma}h_{\nu}^{\mu}+\bar{\nabla}_{\mu}\bar{\nabla}_{\nu}h_{\sigma}^{\mu}-\bar{\Box}h_{\sigma\nu}-\bar{\nabla}_{\sigma}\bar{\nabla}_{\nu}h\right),\label{eq:Linear_Ricci}\end{equation}
 \begin{equation}
R_{L}=\bar{g}^{\alpha\beta}R_{\alpha\beta}^{L}-\bar{R}^{\alpha\beta}h_{\alpha\beta}.\label{eq:Linear_R}\end{equation}
 Note again that the above formulae work for \emph{any} background
space including constant curvature spaces which we shall concentrate
below.

\section{BI-Type Actions at $O\left(h_{\mu\nu}^{2}\right)$}

\subsection{General analysis}

A generic Born-Infeld type action which reproduces the Einstein-Hilbert
theory with a bare cosmological constant ($\Lambda_{0}$) at the first
order in small \emph{curvature} expansion is of the form\begin{equation}
I=\frac{2}{\kappa\alpha}\int d^{D}x\,\left[\sqrt{-\det\left(g_{\mu\nu}+A_{\mu\nu}\right)}-\left(\alpha\Lambda_{0}+1\right)\sqrt{-\det g}\right],\label{eq:Generic_BI_action}\end{equation}
 where $A_{\mu\nu}$ should read as $A_{\mu\nu}=\alpha\left(R_{\mu\nu}+\beta\tilde{R}_{\mu\nu}\right)+O\left(R^{2}\right)$
with the definition $\tilde{R}_{\mu\nu}\equiv R_{\mu\nu}-\frac{1}{D}g_{\mu\nu}R$.
The $O\left(R^{2}\right)$ terms may involve rank $\left(0,2\right)$
combinations of the Riemann and the Ricci tensors, the metric and
the scalar curvature. It could also involve the derivatives of these
tensors, but we will not explicitly consider such actions, and we
will demand parity invariance, so we do not use the $\epsilon^{\mu\nu\lambda\sigma\dots\theta}$
tensor in the construction of $A_{\mu\nu}$. Of course, all these
technical restrictions can be removed and the following discussion
can be extended without much difficulty to cover the type of actions
used in \cite{Gullu3}. Here, the dimensionful parameter $\alpha$
with a $\left(\text{mass}\right)^{-2}$ dimension appears only beyond
the Einstein-Hilbert theory, and $\kappa$ is related to the Newton's
constant. Note that in the Born-Infeld extension of Maxwell's theory,
$\sqrt{-\det\left(g_{\mu\nu}+bF_{\mu\nu}\right)}$, one \emph{must}
introduce a dimensionful parameter $b$, since Maxwell's theory is
scale invariant, but the BI theory cannot be. On the other hand, gravity
is not scale invariant and in principle one need not introduce a new
scale, one can simply use the already existing two scales $\kappa$
and $\Lambda_{0}$. Nevertheless, introducing a new scale $\alpha$
gives more flexibility to the theory.

To study the unitarity of (\ref{eq:Generic_BI_action}), one should
consider the quadratic fluctuations around a critical point of the
action. Assuming that $\bar{g}_{\mu\nu}$ is the critical point and
$h_{\mu\nu}$ is the fluctuation, we should compute the $O\left(h^{2}\right)$
terms in the action. To do this by just pulling out the volume density,
it is convenient to write the action in the form \begin{equation}
I=\frac{2}{\kappa\alpha}\int d^{D}x\,\sqrt{-g}\left[\sqrt{-\det\left(\delta_{\nu}^{\rho}+g^{\rho\mu}A_{\mu\nu}\right)}-\left(\alpha\Lambda_{0}+1\right)\right],\end{equation}
 Using the second order expansion of the inverse metric, (\ref{eq:Oh2_of_inv_met}),
and assuming an expansion of $A_{\mu\nu}$ in the metric perturbation
as\begin{equation}
A_{\mu\nu}\equiv\bar{A}_{\mu\nu}+\tau A_{\mu\nu}^{\left(1\right)}+\tau^{2}A_{\mu\nu}^{\left(2\right)}+O\left(\tau^{3}\right),\end{equation}
 one has\begin{equation}
g^{\rho\mu}A_{\mu\nu}=\bar{g}^{\rho\mu}\bar{A}_{\mu\nu}+\tau\left(\bar{g}^{\rho\mu}A_{\mu\nu}^{\left(1\right)}-h^{\rho\mu}\bar{A}_{\mu\nu}\right)+\tau^{2}\left(\bar{g}^{\rho\mu}A_{\mu\nu}^{\left(2\right)}-h^{\rho\mu}A_{\mu\nu}^{\left(1\right)}+h^{\rho\sigma}h_{\sigma}^{\mu}\bar{A}_{\mu\nu}\right).\label{eq:hmn_exp_of_gA}\end{equation}
 In order to find the second order action in metric perturbation,
let us separate the background part of $g^{\rho\mu}A_{\mu\nu}$ and
define $\tau B_{\nu}^{\rho}\equiv g^{\rho\mu}A_{\mu\nu}-\bar{g}^{\rho\mu}\bar{A}_{\mu\nu}$,
whose introduction will make the expansion more transparent. For a
maximally symmetric constant curvature background, one has $\bar{A}_{\mu\nu}\equiv a\bar{g}_{\mu\nu}$
where $a$ is a dimensionless constant \emph{fixed} in the theory
in terms of the dimensionful parameters such as $\Lambda_{0}$, $\alpha$,
\emph{etc}. The effective cosmological constant $\Lambda$ will also
be fixed by the dimensionful parameters. For complicated actions,
even finding $\Lambda$ is a nontrivial problem. The obvious and the
conventional method is to find the equations of motion and insert
the maximally symmetric solution. But, finding the equations of motion
for these actions is simply too complicated. Therefore, we will give
a method which bypasses this. Then, $B_{\nu}^{\rho}$ becomes\begin{equation}
B_{\nu}^{\rho}=\left(\bar{g}^{\rho\mu}A_{\mu\nu}^{\left(1\right)}-ah_{\nu}^{\rho}\right)+\tau\left(\bar{g}^{\rho\mu}A_{\mu\nu}^{\left(2\right)}-h^{\rho\mu}A_{\mu\nu}^{\left(1\right)}+ah^{\rho\sigma}h_{\sigma\nu}\right).\end{equation}
 Now, we can re-express the BI action with the help of the $B_{\nu}^{\rho}$
tensor\begin{align}
I & =\frac{2}{\kappa\alpha}\int d^{D}x\,\sqrt{-g}\left\{ \sqrt{-\det\left[\left(1+a\right)\delta_{\nu}^{\rho}+\tau B_{\nu}^{\rho}\right]}-\left(\alpha\Lambda_{0}+1\right)\right\} \nonumber \\
 & =\frac{2}{\kappa\alpha}\left(1+a\right)^{\frac{D-4}{2}}\int d^{D}x\,\sqrt{-g}\left\{ \left(1+a\right)^{2}\sqrt{-\det\left[\delta_{\nu}^{\rho}+\frac{\tau}{\left(1+a\right)}B_{\nu}^{\rho}\right]}-\left(1+a\right)^{\frac{4-D}{2}}\left(\alpha\Lambda_{0}+1\right)\right\} ,\label{eq:B_tensor_action}\end{align}
 where $a\ne-1$ which is required in order to have a well-defined
leading order: if this requirement is not put, then the flat space
limit cannot be reproduced in the limit of vanishing cosmological
constant. {[}For example, if one had fixed $\alpha=-\frac{1}{\Lambda_{0}}$
with $A_{\mu\nu}=\alpha R_{\mu\nu}$, then one would not have a proper
flat space limit.{]} Here, the factor $\left(1+a\right)^{2}$ is left
in front of the determinantal part in order not to introduce $a$
factors in the second order terms coming from the expansion of the
determinant. To find the second order expansion of the action in the
metric perturbation, let us Taylor expand the determinant in terms
of traces up to the order that we shall need\begin{align}
\left[\det\left(1+M\right)\right]^{1/2}= & 1+\frac{1}{2}\text{Tr}M+\frac{1}{8}\left(\text{Tr}M\right)^{2}-\frac{1}{4}\text{Tr}\left(M^{2}\right)\nonumber \\
 & +\frac{1}{6}\text{Tr}\left(M^{3}\right)-\frac{1}{8}\text{Tr}\left(M^{2}\right)\text{Tr}M+\frac{1}{48}\left(\text{Tr}M\right)^{3}+O\left(M^{4}\right).\label{eq:Small_M_expansion}\end{align}
 With this formula, the second order expansion of $\sqrt{-g}$ becomes\begin{equation}
\sqrt{-\det g_{\mu\nu}}=\sqrt{-\det\left(\bar{g}_{\mu\nu}+\tau h_{\mu\nu}\right)}=\sqrt{-\bar{g}}\left[1+\frac{\tau}{2}h+\frac{1}{8}\tau^{2}\left(h^{2}-2h_{\mu\nu}^{2}\right)+O\left(\tau^{3}\right)\right].\label{eq:Oh2_exp_of_rootg}\end{equation}
 Then, after using the expansions of the Lagrangian and $\sqrt{-g}$
in (\ref{eq:B_tensor_action}) one obtains up to $O\left(\tau^{3}\right)$\begin{align}
I & =\frac{2}{\kappa\alpha}\left(1+a\right)^{\frac{D-4}{2}}\int d^{D}x\,\sqrt{-\bar{g}}\Biggl\{\left[\left(1+a\right)^{2}-\left(1+a\right)^{\frac{4-D}{2}}\left(\alpha\Lambda_{0}+1\right)\right]\nonumber \\
 & \phantom{=}+\frac{\tau}{2}\left[\left(1+a\right)B_{\rho}^{\rho}+\left[\left(1+a\right)^{2}-\left(1+a\right)^{\frac{4-D}{2}}\left(\alpha\Lambda_{0}+1\right)\right]h\right]\nonumber \\
 & \phantom{=}+\frac{\tau^{2}}{8}\left[\left(B_{\rho}^{\rho}\right)^{2}-2B_{\nu}^{\rho}B_{\rho}^{\nu}+2\left(1+a\right)hB_{\rho}^{\rho}+\left[\left(1+a\right)^{2}-\left(1+a\right)^{\frac{4-D}{2}}\left(\alpha\Lambda_{0}+1\right)\right]\left(h^{2}-2h_{\mu\nu}^{2}\right)\right]\Biggr\}.\label{eq:Expansion_of_B_tensor_act}\end{align}
 $O\left(\tau^{0}\right)$ term just gives the value of the action
for the vacuum solution and it will not be relevant anymore. But,
it gives us some crucial information about the BI-type actions, that
is for even dimensions the value of the constant curvature is not
bounded by the action; however, for odd dimensions $a>-1$ is required
for the reality of the action. Now, we would like to go back to our
original tensor $A_{\mu\nu}$. First, we write the $O\left(\tau\right)$
term in the above expression in terms of $A_{\mu\nu}$. This term
gives the nonlinear equation of motion for the constant curvature
background. One needs to find what the zeroth order of $B_{\rho}^{\rho}$
is in terms of $A_{\mu\nu}$. This is given as \begin{equation}
B_{\rho}^{\rho}=\bar{g}^{\rho\mu}A_{\mu\rho}^{\left(1\right)}-ah+O\left(\tau\right).\end{equation}
 Then, the action at the first order reads\begin{equation}
I_{O\left(h\right)}=\frac{\left(1+a\right)^{\frac{D-4}{2}}}{\kappa\alpha}\int d^{D}x\,\sqrt{-\bar{g}}\left[\left(1+a\right)\left(\bar{g}^{\rho\mu}A_{\mu\rho}^{\left(1\right)}+h\right)-\left(1+a\right)^{\frac{4-D}{2}}\left(\alpha\Lambda_{0}+1\right)h\right].\label{eq:Oh_action}\end{equation}
 After removing possible boundary terms, taking the variation with
respect to $h_{\mu\nu}$ or more concisely looking at the coefficient
of $h^{\mu\nu}$ and equating it to zero yields the source-free nonlinear
equation of motion for a constant curvature background, namely, the
equation of motion that relates $\Lambda$ to $\Lambda_{0}$ and the
other parameters of the theory. Hence, to get the vacuum of the theory,
one need not explicitly find the equations of motion which is straightforward
but quite tedious. 

Now, let us find the quadratic action in $h_{\mu\nu}$ in terms of
$A_{\mu\nu}$. The $\left(B_{\rho}^{\rho}\right)^{2}-2B_{\nu}^{\rho}B_{\rho}^{\nu}+2\left(1+a\right)hB_{\rho}^{\rho}$
terms in (\ref{eq:Expansion_of_B_tensor_act}) can be written in terms
of $A_{\mu\nu}$ as\begin{align}
\left(B_{\rho}^{\rho}\right)^{2}-2B_{\nu}^{\rho}B_{\rho}^{\nu}+2\left(1+a\right)hB_{\rho}^{\rho}= & \left(\bar{g}^{\mu\nu}A_{\mu\nu}^{\left(1\right)}\right)^{2}-2A_{\mu\nu}^{\left(1\right)}A_{\left(1\right)}^{\mu\nu}\nonumber \\
 & +h^{\mu\nu}\left[4aA_{\mu\nu}^{\left(1\right)}+2\bar{g}_{\mu\nu}\bar{g}^{\rho\sigma}A_{\rho\sigma}^{\left(1\right)}-2a^{2}h_{\mu\nu}-a\left(2+a\right)\bar{g}_{\mu\nu}h\right].\end{align}
 Contribution coming from the $\tau\left(1+a\right)B_{\rho}^{\rho}$
term in (\ref{eq:Expansion_of_B_tensor_act}) is\begin{equation}
B_{\rho}^{\rho}=O\left(\tau^{0}\right)+\tau\left[\bar{g}^{\mu\nu}A_{\mu\nu}^{\left(2\right)}-h^{\mu\nu}\left(A_{\mu\nu}^{\left(1\right)}-ah_{\mu\nu}\right)\right].\end{equation}
 In all together, the quadratic action in $h_{\mu\nu}$ in terms of
$A_{\mu\nu}$ boils down to\begin{align}
I_{O\left(h^{2}\right)}= & -\frac{\left(1+a\right)^{\frac{D-4}{2}}}{\kappa\alpha}\int d^{D}x\,\sqrt{-\bar{g}}\left\{ \frac{1}{2}A_{\mu\nu}^{\left(1\right)}A_{\left(1\right)}^{\mu\nu}-\frac{1}{4}\left(\bar{g}^{\mu\nu}A_{\mu\nu}^{\left(1\right)}\right)^{2}-\left(1+a\right)\bar{g}^{\mu\nu}A_{\mu\nu}^{\left(2\right)}\right.\nonumber \\
 & \left.+h^{\mu\nu}\left(A_{\mu\nu}^{\left(1\right)}-\frac{1}{2}\bar{g}_{\mu\nu}\bar{g}^{\rho\sigma}A_{\rho\sigma}^{\left(1\right)}\right)-\frac{1}{4}\left[1-\left(1+a\right)^{\frac{4-D}{2}}\left(\alpha\Lambda_{0}+1\right)\right]\left(h^{2}-2h_{\mu\nu}^{2}\right)\right\} .\label{eq:Oh2_action}\end{align}
 To remove a possible confusion coming from the notation, we should
note what is represented by the term $A_{\mu\nu}^{\left(1\right)}A_{\left(1\right)}^{\mu\nu}$:
It is basically $A_{\mu\nu}^{\left(1\right)}A_{\left(1\right)}^{\mu\nu}\equiv\bar{g}^{\mu\alpha}\bar{g}^{\nu\beta}A_{\mu\nu}^{\left(1\right)}A_{\alpha\beta}^{\left(1\right)}$,
that is $A_{\left(1\right)}^{\mu\nu}$ \emph{does not} represent the
first order of $A^{\mu\nu}$. If required, we show the first order
of $A^{\mu\nu}$ as $\left(A^{\mu\nu}\right)_{\left(1\right)}$. Equation
(\ref{eq:Oh2_action}) is our main formula which can be applied to
any BI-type action for any value of the constant curvature {[}i.e.
we have not done a small curvature expansion, that is, the formula
at $O\left(h^{2}\right)$ takes care of all the contributions coming
from all powers of the curvature{]}. Let us summarize what one needs
to do to analyze the unitarity of a given BI gravity: One computes
$A_{\mu\nu}^{\left(1\right)}$ and $A_{\mu\nu}^{\left(2\right)}$,
and using (\ref{eq:Oh_action}) one finds the vacuum of the theory,
and finally computes the $O\left(h^{2}\right)$ action via (\ref{eq:Oh2_action}).
Then, this action can be studied using conventional techniques that
were discussed in \cite{Gullu1}. Of course, as we shall see below
with some examples, depending on the complexity of $A_{\mu\nu}$,
explicit computation of (\ref{eq:Oh2_action}) could be a very cumbersome
problem in generic dimensions. But, a close scrutiny of it reveals
remarkable simplifications in even dimensions, higher than two, and
especially in four dimensions. Such simplifications, in four dimensions,
will provide us with another method of analyzing the unitarity of
the BI gravities, namely, the method of Hindawi \emph{et al} \cite{Hindawi}
that leads to the construction of an equivalent quadratic action (in
curvature) whose unitarity has been already studied by conventional
methods. Let us concentrate on $D=4$ first whose action is \begin{align}
I_{O\left(h^{2}\right)}= & -\frac{1}{\kappa\alpha}\int d^{4}x\,\sqrt{-\bar{g}}\left\{ \frac{1}{2}A_{\mu\nu}^{\left(1\right)}A_{\left(1\right)}^{\mu\nu}-\frac{1}{4}\left(\bar{g}^{\mu\nu}A_{\mu\nu}^{\left(1\right)}\right)^{2}-\left(1+a\right)\bar{g}^{\mu\nu}A_{\mu\nu}^{\left(2\right)}\right.\nonumber \\
 & \left.+h^{\mu\nu}\left(A_{\mu\nu}^{\left(1\right)}-\frac{1}{2}\bar{g}_{\mu\nu}\bar{g}^{\rho\sigma}A_{\rho\sigma}^{\left(1\right)}\right)+\frac{1}{4}\alpha\Lambda_{0}\left(h^{2}-2h_{\mu\nu}^{2}\right)\right\} .\label{eq:Oh2_action_4D}\end{align}
 By examining this action, one can figure out an interesting relation
between the metric perturbation expansion that led to this action
and the $A_{\mu\nu}$ expansion of (\ref{eq:Generic_BI_action}).
Remember that $A_{\mu\nu}$ is dimensionless, so assuming proper convergence,
a Taylor series expansion over $A_{\mu\nu}$ is legitimate. If $A_{\mu\nu}$
involves terms of $O\left(R^{2}\right)$ and/or any other higher curvature
terms, the $A_{\mu\nu}$ expansion is not simply equal to the curvature
expansion in which the expansion is over the nondimensional quantity
$\alpha R$. Let us write symbolically the expansion of (\ref{eq:Generic_BI_action})
in $A_{\mu\nu}$ as\begin{multline}
I=\frac{2}{\kappa\alpha}\int d^{4}x\,\left[\sqrt{-\det\left(g_{\mu\nu}+A_{\mu\nu}\right)}-\left(\alpha\Lambda_{0}+1\right)\sqrt{-\det g}\right]\\
\sim\frac{2}{\kappa\alpha}\int d^{4}x\,\sqrt{-g}\left[\sum_{n=0}^{\infty}c_{n}A^{n}-\left(\alpha\Lambda_{0}+1\right)\right]=\int d^{4}x\,\sqrt{-g}\left[\frac{1}{\kappa}\left(R-2\Lambda_{0}\right)+\frac{2}{\kappa\alpha}\sum_{n=2}^{\infty}c_{n}A^{n}\right],\label{eq:A_exp_of_general_BI}\end{multline}
 where the last equality follows from our assumption that Einstein-Hilbert
action is reproduced at the lowest order. Note that, up to $n=3$,
this expansion can be obtained with help of (\ref{eq:Small_M_expansion}),
and the $n^{\text{th}}$ order term represented with $A^{n}$ involves
terms like $A^{n}$, $A^{n-2}A_{\mu\nu}^{2}$, $A^{n-3}A_{\rho}^{\mu}A_{\nu}^{\rho}A_{\mu}^{\nu}$,
\emph{etc}. In principle, each order in (\ref{eq:A_exp_of_general_BI})
contributes to the quadratic action in the metric perturbation given
in (\ref{eq:Oh2_action_4D}), but we will see that this is not the
case in four dimensions. The $O\left(h^{2}\right)$ contributions
coming from the $O\left(A^{n}\right)$ terms where $n\ge2$ have the
form\begin{align}
I_{O\left(h^{2}\right)}^{\left(n\right)}= & \int d^{4}x\,\sqrt{-\bar{g}}c_{n}\Biggl\{\bar{A}^{n-2}\left[d_{n1}A_{\mu\nu}^{\left(1\right)}A_{\left(1\right)}^{\mu\nu}+d_{n2}\left(\bar{g}^{\mu\nu}A_{\mu\nu}^{\left(1\right)}\right)^{2}\right]\nonumber \\
 & +\bar{A}^{n-1}\left[d_{n3}A_{\mu\nu}^{\left(2\right)}+d_{n4}h^{\mu\nu}A_{\mu\nu}^{\left(1\right)}+d_{n5}h\bar{g}^{\rho\sigma}A_{\rho\sigma}^{\left(1\right)}\right]+\bar{A}^{n}\left[d_{n6}h_{\mu\nu}^{2}+d_{n7}h^{2}\right]\Biggr\},\label{eq:Oh2_of_An}\end{align}
 where $\bar{A}$ is defined as $\bar{A}_{\mu\nu}\equiv a\bar{g}_{\mu\nu}$
as above, and the coefficients $d_{n}$ are just numbers. Therefore,
the $O\left(h^{2}\right)$ contributions coming from the $O\left(A^{n}\right)$
terms are in the form of $\left[e_{n2}\left(h\right)a^{n-2}+e_{n1}\left(h\right)a^{n-1}+e_{n0}\left(h\right)a^{n}\right]$.
Hence, one expects that if each order in the $A_{\mu\nu}$ expansion
of (\ref{eq:Generic_BI_action}) contributes to the quadratic action
in metric fluctuations, then that action will be composed of the seven
terms specified in (\ref{eq:Oh2_of_An}) with a coefficient which
is a power series in $a$. With this result, one can trace the contribution
coming from each order in (\ref{eq:A_exp_of_general_BI}) to the $O\left(h^{2}\right)$
action (\ref{eq:Oh2_action_4D}). Let us investigate each term in
(\ref{eq:Oh2_action_4D}) in order to find which orders in the $A_{\mu\nu}$
expansion contributes. The first two terms in (\ref{eq:Oh2_action_4D}),
which are quadratic in $A_{\mu\nu}$, have coefficients that do not
depend on $a$. Therefore, these two terms involve $O\left(h^{2}\right)$
contributions only coming from the second order terms in the $A_{\mu\nu}$
expansion of (\ref{eq:Generic_BI_action}). The coefficient of the
third term in (\ref{eq:Oh2_action_4D}) is $\left(1+a\right)$, so
it is composed of contributions coming from $O\left(A\right)$ and
$O\left(A^{2}\right)$ terms in the $A_{\mu\nu}$ expansion (\ref{eq:A_exp_of_general_BI}).
The fourth term has a coefficient which does not depend on $a$, so
it comes from the first order of the $A_{\mu\nu}$ expansion. Thus,
all the $O\left(h^{2}\right)$ contributions coming from $O\left(A^{n}\right)$
terms with $n>2$ are identically zero for four-dimensional BI-type
actions, and as we will see this curious case has a generalization
to higher even dimensions. With these observations, one can deduce
the fact that in four dimensions (\ref{eq:Oh2_action_4D}) can be
obtained first by making an expansion in $A_{\mu\nu}$ up to third
order via (\ref{eq:Small_M_expansion}), and then by finding the quadratic
action in metric fluctuations. In other words, remarkably the free
theory of the following actions are exactly the same:\begin{equation}
I=\frac{2}{\kappa\alpha}\int d^{4}x\,\left[\sqrt{-\det\left(g_{\mu\nu}+A_{\mu\nu}\right)}-\left(\alpha\Lambda_{0}+1\right)\sqrt{-\det g}\right],\label{eq:General_BI_again}\end{equation}
and\begin{align}
I_{O\left(A^{2}\right)} & =\frac{2}{\kappa\alpha}\int d^{4}x\,\sqrt{-g}\left[\frac{1}{2}g^{\mu\nu}A_{\mu\nu}-\alpha\Lambda_{0}+\frac{1}{8}\left(g^{\mu\nu}A_{\mu\nu}\right)^{2}-\frac{1}{4}A_{\mu\nu}^{2}\right],\label{eq:OA2_in_4D}\end{align}
 which was obtained by expanding (\ref{eq:General_BI_again}). Here,
note that we truncated the $A_{\mu\nu}$ expansion at the second order,
\emph{but} \emph{we do not require $A_{\mu\nu}$ to be small}. This
truncation can be done and the equality of the above two actions at
the free level can be achieved merely due to the fact that contributions
of the higher order terms in the $A_{\mu\nu}$ expansion to the quadratic
action in metric fluctuations are identically zero. Such a remarkable
cancellation in four dimensions is related to the fact that we have
the square root of the determinant of a linear combination of matrix
functions one of which is expanded around a constant curvature space
and it would not work for a generic background. Let us verify this
result by explicitly calculating the quadratic action in metric fluctuations
for (\ref{eq:OA2_in_4D}). However, to be as general as possible and
to see some cancellations, let us work in $D$ dimensions where only
measure in (\ref{eq:OA2_in_4D}) changes to $d^{D}x$. Then, expanding
each term in (\ref{eq:OA2_in_4D}) by using (\ref{eq:Oh2_exp_of_rootg})
and (\ref{eq:hmn_exp_of_gA}) with $\bar{A}_{\mu\nu}\equiv a\bar{g}_{\mu\nu}$
one has\begin{equation}
g^{\mu\nu}A_{\mu\nu}=aD+\tau\left(\bar{g}^{\mu\nu}A_{\mu\nu}^{\left(1\right)}-ah\right)+\tau^{2}\left(\bar{g}^{\mu\nu}A_{\mu\nu}^{\left(2\right)}-h^{\mu\nu}A_{\mu\nu}^{\left(1\right)}+ah_{\mu\nu}^{2}\right),\end{equation}
 and all together up to quadratic order, the action reads\begin{align}
I_{O\left(A^{2}\right)} & =\frac{1}{\kappa\alpha}\int d^{D}x\,\sqrt{-\bar{g}}\Biggl\{\left[aD-\frac{a^{2}D}{2}+\frac{a^{2}D^{2}}{4}-\alpha\Lambda_{0}\right]\nonumber \\
 & \phantom{=\frac{1}{\kappa\alpha}\int d^{D}x\,\sqrt{-\bar{g}}}+\tau\left[\left(1+\frac{aD}{2}-a\right)\bar{g}^{\mu\nu}A_{\mu\nu}^{\left(1\right)}+\frac{a\left(D-2\right)}{2}\left(1+\frac{\left(D-4\right)a}{4}\right)h-\alpha\Lambda_{0}h\right]\nonumber \\
 & \phantom{=\frac{1}{\kappa\alpha}\int d^{D}x\,\sqrt{-\bar{g}}}-\tau^{2}\biggl[\frac{1}{2}A_{\left(1\right)}^{\mu\nu}A_{\mu\nu}^{\left(1\right)}-\frac{1}{4}\left(\bar{g}^{\mu\nu}A_{\mu\nu}^{\left(1\right)}\right)^{2}-\left(1+\frac{aD}{2}-a\right)\bar{g}^{\mu\nu}A_{\mu\nu}^{\left(2\right)}\label{eq:h_exp_of_OA2}\\
 & \phantom{=\frac{1}{\kappa\alpha}\int d^{D}x\,\sqrt{-\bar{g}}-\tau^{2}}+\left(1+\frac{aD}{2}-2a\right)h^{\mu\nu}\left(A_{\mu\nu}^{\left(1\right)}-\frac{1}{2}\bar{g}_{\mu\nu}\bar{g}^{\rho\sigma}A_{\rho\sigma}^{\left(1\right)}\right)\nonumber \\
 & \phantom{=\frac{1}{\kappa\alpha}\int d^{D}x\,\sqrt{-\bar{g}}-\tau^{2}}-\frac{\left(D-4\right)}{8}\left(a+\frac{D-6}{4}a^{2}\right)\left(h^{2}-2h_{\mu\nu}^{2}\right)+\frac{\alpha\Lambda_{0}}{4}\left(h^{2}-2h_{\mu\nu}^{2}\right)\biggr]\Biggr\}.\nonumber \end{align}
 In obtaining this result, one should rewrite $\left(A^{\mu\nu}\right)_{\left(2\right)}$
and $\left(A^{\mu\nu}\right)_{\left(1\right)}$ coming from $A_{\mu\nu}^{2}$
in (\ref{eq:OA2_in_4D}) in terms of $A_{\mu\nu}^{\left(2\right)}$
and $A_{\mu\nu}^{\left(1\right)}$ as\begin{equation}
\left(A^{\mu\nu}\right)_{\left(2\right)}=\left(g^{\mu\alpha}g^{\nu\beta}A_{\alpha\beta}\right)_{\left(2\right)}=\bar{g}^{\mu\alpha}\bar{g}^{\nu\beta}A_{\alpha\beta}^{\left(2\right)}+3ah_{\rho}^{\mu}h^{\rho\nu}-\bar{g}^{\mu\alpha}h^{\nu\beta}A_{\alpha\beta}^{\left(1\right)}-\bar{g}^{\nu\beta}h^{\mu\alpha}A_{\alpha\beta}^{\left(1\right)},\end{equation}
 \begin{equation}
\left(A^{\mu\nu}\right)_{\left(1\right)}=\left(g^{\mu\alpha}g^{\nu\beta}A_{\alpha\beta}\right)^{\left(1\right)}=\bar{g}^{\mu\alpha}\bar{g}^{\nu\beta}A_{\alpha\beta}^{\left(1\right)}-2ah^{\mu\nu}.\end{equation}
 Let us now concentrate only on the $O\left(\tau^{2}\right)$ terms:\begin{align}
I_{O\left(h^{2}\right)}^{O\left(A^{2}\right)} & =-\frac{1}{\kappa\alpha}\int d^{D}x\,\sqrt{-\bar{g}}\left\{ \frac{1}{2}A_{\left(1\right)}^{\mu\nu}A_{\mu\nu}^{\left(1\right)}-\frac{1}{4}\left(\bar{g}^{\mu\nu}A_{\mu\nu}^{\left(1\right)}\right)^{2}-\left(1+\frac{aD}{2}-a\right)\bar{g}^{\mu\nu}A_{\mu\nu}^{\left(2\right)}\right.\nonumber \\
 & \phantom{=-\frac{1}{\kappa\alpha}\int d^{D}x\,\sqrt{-\bar{g}}}\left.+\left(1+\frac{aD}{2}-2a\right)h^{\mu\nu}\left(A_{\mu\nu}^{\left(1\right)}-\frac{1}{2}\bar{g}_{\mu\nu}\bar{g}^{\rho\sigma}A_{\rho\sigma}^{\left(1\right)}\right)\right.\label{eq:Oh2_of_OA2_for_any_D}\\
 & \phantom{=-\frac{1}{\kappa\alpha}\int d^{D}x\,\sqrt{-\bar{g}}}\left.-\frac{\left(D-4\right)}{8}\left[a+\frac{\left(D-6\right)a^{2}}{4}\right]\left(h^{2}-2h_{\mu\nu}^{2}\right)+\frac{\alpha\Lambda_{0}}{4}\left(h^{2}-2h_{\mu\nu}^{2}\right)\right\} .\nonumber \end{align}
 In four dimensions, (\ref{eq:Oh2_of_OA2_for_any_D}) reduces to (\ref{eq:Oh2_action_4D})
as it was promised. In Appendix A, we give a simple example with two-dimensional
matrix functions that shows the connection between the $A_{\mu\nu}$
expansion and the metric perturbation expansion. In generic even dimensions
with $D=2n+2$, if one wants to carry out a similar analysis, then
one has to expand up to $O\left(A^{n+1}\right)$ with $n\ge1$. But,
again we should stress that the compact formula (\ref{eq:Oh2_action})
works all the time without recourse to such an expansion. However,
depending on the complexity of $A_{\mu\nu}$, one can choose to use
either the expansion method or the compact expression. As for odd
dimensions, because of the nonpolynomial prefactor $\left(1+a\right)^{\frac{D-4}{2}}$
in (\ref{eq:Oh2_action}), all the terms in the $A_{\mu\nu}$ expansion
(or the small curvature expansion) contribute. The most efficient
way to get the quadratic fluctuations for odd dimensions is to use
(\ref{eq:Oh2_action}).

A similar analysis can be done for the $O\left(h\right)$ action in
four dimension which is \begin{equation}
I_{O\left(h\right)}=\frac{1}{\kappa\alpha}\int d^{4}x\,\sqrt{-\bar{g}}\left[\left(1+a\right)\bar{g}^{\rho\mu}A_{\mu\rho}^{\left(1\right)}+\left(a-\alpha\Lambda_{0}\right)h\right].\label{eq:Oh_action_4D}\end{equation}
 This action involves contributions coming only from the second order
expansion of (\ref{eq:Generic_BI_action}) in $A_{\mu\nu}$ just as
the $O\left(h^{2}\right)$ action. In order to understand this behavior,
let us first look at the $O\left(h\right)$ contributions coming from
the $O\left(A^{n}\right)$ term for $n\ge2$\begin{align}
I_{O\left(h\right)}^{\left(n\right)}=\int d^{4}x\,\sqrt{-\bar{g}}c_{n} & \left[\bar{A}^{n-1}d_{n1}\left(\bar{g}^{\mu\nu}A_{\mu\nu}^{\left(1\right)}\right)+d_{n2}\bar{A}^{n}h\right],\label{eq:Oh_of_An}\end{align}
 where $d_{n}$ coefficients are just numbers. Therefore, the $O\left(h\right)$
contributions coming from the $O\left(A^{n}\right)$ terms are simply
in the form of $\left[e_{n1}\left(h\right)a^{n-1}+e_{n0}\left(h\right)a^{n}\right]$,
since $\bar{A}\sim a$. Hence, one expects that if each order in the
$A_{\mu\nu}$ expansion of (\ref{eq:Generic_BI_action}) contributes
to the linear action in $h_{\mu\nu}$, then it will be composed of
the two terms specified in (\ref{eq:Oh_of_An}) with coefficients
that are of the form $a^{n}$. With this result, one can trace the
contribution coming from each order in (\ref{eq:A_exp_of_general_BI})
to the $O\left(h\right)$ action (\ref{eq:Oh_action_4D}). Let us
investigate each term in (\ref{eq:Oh_action_4D}) in order to find
which orders in the $A_{\mu\nu}$ expansion contribute. The first
term in (\ref{eq:Oh_action_4D}) has a coefficient of $\left(1+a\right)$.
Therefore, this term involves $O\left(h\right)$ contributions coming
from the second order terms in the $A_{\mu\nu}$ expansion of (\ref{eq:Generic_BI_action}).
The coefficient of $h$ in (\ref{eq:Oh_action_4D}) is also first
order in $a$, but this time it implies that only the first order
of $A_{\mu\nu}$ expansion contributes. Therefore, the vacuum of (\ref{eq:Generic_BI_action})
and (\ref{eq:OA2_in_4D}) are the same. One can verify this result
explicitly from $O\left(h\right)$ of the $O\left(A_{\mu\nu}^{2}\right)$
action which can be read from (\ref{eq:h_exp_of_OA2}) as\[
I_{O\left(h\right)}^{O\left(A^{2}\right)}=\frac{1}{\kappa\alpha}\int d^{D}x\,\sqrt{-\bar{g}}\left[\left(1+\frac{aD}{2}-a\right)\bar{g}^{\mu\nu}A_{\mu\nu}^{\left(1\right)}+\frac{a\left(D-2\right)}{2}\left(1+\frac{\left(D-4\right)a}{4}\right)h-\alpha\Lambda_{0}h\right].\]
 This action reduces to (\ref{eq:Oh_action_4D}) in four dimensions.
Just like the analysis of $O\left(h^{2}\right)$, for generic even
dimensions $D=2n+2$ one has to expand (\ref{eq:Generic_BI_action})
to $O\left(A^{n+1}\right)$ with $n\ge1$, then find the vacuum of
the theory. For odd dimensions, since all the powers of $A^{n}$ contribute,
the most efficient way to find the vacuum of the theory is to use
(\ref{eq:Oh_action}).

\subsection{An example}

To apply our tools, for the sake of simplicity, let us consider the
following model which we know to be nonunitary even around the flat
space: \begin{equation}
I=\frac{2}{\kappa\alpha}\int d^{4}x\,\left[\sqrt{-\det\left(g_{\mu\nu}+\alpha R_{\mu\nu}\right)}-\left(\alpha\Lambda_{0}+1\right)\sqrt{-\det g}\right].\label{eq:Rmn_action}\end{equation}
 Here, according to our results above, one expects (which we shall
verify below with several different techniques) that the second order
action in the metric perturbation $h_{\mu\nu}$ involves contributions
only coming from the $O\left[\left(\alpha R\right)^{2}\right]$ expansion:\begin{equation}
I_{O\left(R^{2}\right)}=\frac{2}{\kappa\alpha}\int d^{4}x\sqrt{-g}\,\left[\frac{\alpha}{2}\left(R-2\Lambda_{0}\right)-\frac{\alpha^{2}}{4}\left(R_{\mu\nu}^{2}-\frac{1}{2}R^{2}\right)\right].\label{eq:OR2_of_Rmn_action}\end{equation}
 Therefore, the $O\left[\left(\alpha R\right)^{3}\right]$, $O\left[\left(\alpha R\right)^{4}\right]$
and etc terms should vanish at $O\left(h^{2}\right)$. Hence, at $O\left(h\right)$
and $O\left(h^{2}\right)$ (\ref{eq:Rmn_action}) and (\ref{eq:OR2_of_Rmn_action})
are equivalent. Let us explicitly show this by analyzing the linearized
free theory of (\ref{eq:Rmn_action}) around the extremum of it by
using (\ref{eq:Oh2_action_4D}).

\subsubsection{Analyzing the BI action formed by the Ricci tensor via second order
perturbations in $h_{\mu\nu}$}

Let us define $A_{\mu\nu}\equiv\alpha R_{\mu\nu}$. Then, $\bar{A}_{\mu\nu}=\alpha\Lambda\bar{g}_{\mu\nu}\Rightarrow a\equiv\alpha\Lambda$
, where $\Lambda$ will be determined in terms of $\Lambda_{0}$.
Then, $A_{\mu\nu}^{\left(1\right)}$ is given as \begin{equation}
A_{\mu\nu}^{\left(1\right)}=\alpha R_{\mu\nu}^{L},\end{equation}
 and $A_{\mu\nu}^{\left(2\right)}=\alpha R_{\nu\sigma}^{\left(2\right)}$
and referring the details to Appendix B, we have\begin{equation}
\alpha\bar{g}^{\mu\nu}R_{\mu\nu}^{\left(2\right)}=\alpha h^{\mu\nu}\left(\frac{1}{2}R_{\mu\nu}^{L}-\frac{1}{4}\bar{g}_{\mu\nu}R_{L}-\frac{\Lambda}{4}\bar{g}_{\mu\nu}h\right).\end{equation}
 First of all, let us determine the nonlinear equations of motion
for the constant curvature background which will relate $\Lambda$
to $\Lambda_{0}$ by using (\ref{eq:Oh_action}) \begin{align}
I_{O\left(h\right)} & =\frac{1}{\kappa\alpha}\int d^{4}x\,\sqrt{-\bar{g}}\left[\left(1+a\right)\left(\bar{g}^{\rho\mu}A_{\mu\rho}^{\left(1\right)}+h\right)-\left(\alpha\Lambda_{0}+1\right)h\right]\nonumber \\
 & =\frac{1}{\kappa\alpha}\int d^{4}x\,\sqrt{-\bar{g}}\left[\alpha\left(1+\alpha\Lambda\right)\bar{\nabla}_{\mu}\left(\bar{\nabla}_{\nu}h^{\mu\nu}-\bar{\nabla}^{\mu}h\right)+\alpha\left(\Lambda-\Lambda_{0}\right)h\right].\end{align}
 Note that this first order correction should be zero around the extremum,
therefore, after dropping the first term which is a boundary term,
one has $\Lambda=\Lambda_{0}$. As for the second order action, one
has (\ref{eq:Oh2_action_4D})\begin{align}
I_{O\left(h^{2}\right)}=-\frac{1}{\kappa\alpha}\int d^{4}x\,\sqrt{-\bar{g}} & \Biggl\{\frac{\alpha^{2}}{2}R_{\mu\nu}^{L}R_{L}^{\mu\nu}-\frac{\alpha^{2}}{4}\left(R_{L}+\Lambda h\right)^{2}\nonumber \\
 & -\left(\alpha+\alpha^{2}\Lambda\right)h^{\mu\nu}\left(\frac{1}{2}R_{\mu\nu}^{L}-\frac{1}{4}\bar{g}_{\mu\nu}R_{L}-\frac{\Lambda}{4}\bar{g}_{\mu\nu}h\right)\\
 & +\alpha h^{\mu\nu}\left[R_{\mu\nu}^{L}-\frac{1}{2}\bar{g}_{\mu\nu}\left(R_{L}+\Lambda h\right)\right]+\frac{\alpha}{4}\Lambda_{0}\left(h^{2}-2h_{\mu\nu}^{2}\right)\Biggr\},\nonumber \end{align}
 where $\int d^{4}x\,\sqrt{-\bar{g}}R_{\mu\nu}^{L}R_{L}^{\mu\nu}$
is calculated in Appendix B as\begin{align}
\int d^{4}x\,\sqrt{-\bar{g}}R_{L}^{\mu\nu}R_{\mu\nu}^{L}=-\frac{1}{2}\int d^{4}x\,\sqrt{-\bar{g}}h^{\mu\nu} & \left[\left(\bar{g}_{\mu\nu}\bar{\Box}-\bar{\nabla}_{\mu}\bar{\nabla}_{\nu}+\Lambda\bar{g}_{\mu\nu}\right)R_{L}+\left(\bar{\Box}\mathcal{G}_{\mu\nu}^{L}-\frac{2\Lambda}{3}\bar{g}_{\mu\nu}R_{L}\right)\right.\nonumber \\
 & \left.-\frac{14\Lambda}{3}R_{\mu\nu}^{L}+\frac{\Lambda}{3}\bar{g}_{\mu\nu}R_{L}+\frac{8\Lambda^{2}}{3}h_{\mu\nu}\right].\end{align}
 Then, after some algebra the quadratic action reduces to \begin{align}
I_{O\left(h^{2}\right)}=-\frac{1}{\alpha\kappa}\int d^{4}x\,\sqrt{-\bar{g}} & \Biggl\{ h^{\mu\nu}\left[\left(\frac{\alpha}{2}+\frac{2\alpha^{2}\Lambda}{3}\right)\mathcal{G}_{\mu\nu}^{L}-\frac{\alpha^{2}}{4}\left(\bar{\Box}\mathcal{G}_{\mu\nu}^{L}-\frac{2\Lambda}{3}\bar{g}_{\mu\nu}R_{L}\right)\right]\nonumber \\
 & -\frac{\alpha}{4}\left(\Lambda-\Lambda_{0}\right)\left(h^{2}-2h_{\mu\nu}^{2}\right)\Biggr\},\end{align}
 where we kept the background gauge noninvariant term (the last part)
just to show an intermediate step of the computation. Once $\Lambda=\Lambda_{0}$
is used, one ends up with \begin{align}
I_{O\left(h^{2}\right)}=-\frac{1}{\alpha\kappa}\int d^{4}x\,\sqrt{-\bar{g}} & h^{\mu\nu}\left[\left(\frac{\alpha}{2}+\frac{2\alpha^{2}\Lambda_{0}}{3}\right)\mathcal{G}_{\mu\nu}^{L}-\frac{\alpha^{2}}{4}\left(\bar{\Box}\mathcal{G}_{\mu\nu}^{L}-\frac{2\Lambda_{0}}{3}\bar{g}_{\mu\nu}R_{L}\right)\right].\label{eq:Oh2_of_Rmn_action}\end{align}
 This action is exactly equivalent to the linearized action one obtains
from the $O\left[\left(\alpha R\right)^{2}\right]$ action (\ref{eq:OR2_of_Rmn_action}).
{[}Note that the linearized version of (\ref{eq:OR2_of_Rmn_action})
has been worked out in several places \cite{DeserTekin,Gullu1}, and
we also reproduce it below.{]} The fact that (\ref{eq:Oh2_of_Rmn_action})
has at most $\alpha^{2}$ terms show that the contributions coming
from all $O\left[\left(\alpha R\right)^{n+2}\right]$ vanish. We stress
once again that this is a highly nontrivial cancellation brought by
the determinantal structure of the action. It is worth to study explicitly
how this cancellation takes place at $O\left[\left(\alpha R\right)^{3}\right]$
which we do now. At this order the action reads\begin{equation}
I_{O\left(R^{3}\right)}=\frac{2}{\kappa\alpha}\int d^{4}x\sqrt{-g}\,\left[\frac{\alpha}{2}\left(R-2\Lambda_{0}\right)-\frac{\alpha^{2}}{4}\left(R_{\mu\nu}^{2}-\frac{1}{2}R^{2}\right)+\frac{\alpha^{3}}{48}\left(8R_{\mu\rho}R_{\nu}^{\rho}R^{\mu\nu}-6R_{\mu\nu}^{2}R+R^{3}\right)\right],\label{eq:OR3_of_Rmn_action}\end{equation}
 and defining \begin{equation}
K\equiv R_{\mu\nu}^{2}-\frac{1}{2}R^{2},\qquad S\equiv8R^{\mu\nu}R_{\mu\alpha}R_{\phantom{\alpha}\nu}^{\alpha}-6RR_{\mu\nu}^{2}+R^{3},\end{equation}
 one has\begin{equation}
I_{O\left(R^{3}\right)}=\frac{1}{\kappa}\int d^{4}x\sqrt{-g}\,\left[\left(R-2\Lambda_{0}\right)-\frac{\alpha}{2}K+\frac{\alpha^{2}}{24}S\right].\end{equation}
 Finding the $O\left(h^{2}\right)$ action of this theory is a very
cumbersome problem. To somewhat simplify this, one can first find
the equations of motion then linearize the equations of motion and
then do the reverse calculus of variations procedure to get the action.
Of course in this process boundary terms are dropped and one has to
be careful with an overall sign that can be fixed by coupling the
gravity action to matter. The equations of motion follow as \begin{align}
\frac{\kappa\alpha}{4}\tau_{\mu\nu} & =-\frac{\alpha}{4}\left[\left(R-2\Lambda_{0}\right)-\frac{\alpha}{2}K+\frac{\alpha^{2}}{24}S\right]g_{\mu\nu}+\frac{\alpha}{2}R_{\mu\nu}\nonumber \\
 & \phantom{=}+\frac{\alpha^{2}}{4}\left[RR_{\mu\nu}-2R_{\lambda\nu\alpha\mu}R^{\lambda\alpha}-\Box\left(R_{\mu\nu}-\frac{1}{2}g_{\mu\nu}R\right)\right]\nonumber \\
 & \phantom{=}+\frac{\alpha^{3}}{4}\left(2R_{\mu}^{\rho}R_{\rho\alpha}R_{\nu}^{\alpha}+\left[g_{\mu\nu}\nabla_{\alpha}\nabla_{\beta}\left(R^{\beta\rho}R_{\rho}^{\alpha}\right)+\Box\left(R_{\nu}^{\rho}R_{\mu\rho}\right)-2\nabla_{\alpha}\nabla_{\mu}\left(R_{\nu}^{\rho}R_{\rho}^{\alpha}\right)\right]\right)\label{eq:EOM_of_OR3_of_Rmn_act}\\
 & \phantom{=}+\frac{\alpha^{3}}{8}\left(\left[2\nabla_{\alpha}\nabla_{\mu}\left(RR_{\nu}^{\alpha}\right)-g_{\mu\nu}\nabla_{\alpha}\nabla_{\beta}\left(RR^{\alpha\beta}\right)-\Box\left(RR_{\mu\nu}\right)\right]-2RR_{\nu}^{\rho}R_{\mu\rho}\right)\nonumber \\
 & \phantom{=}-\frac{\alpha^{3}}{8}\left[\left(g_{\mu\nu}\Box-\nabla_{\nu}\nabla_{\mu}\right)+R_{\mu\nu}\right]\left(R_{\alpha\beta}^{2}-\frac{1}{2}R^{2}\right),\nonumber \end{align}
 where we defined the energy-momentum tensor as $\tau_{\mu\nu}\equiv-\frac{2}{\sqrt{-g}}\frac{\delta I_{\text{matter}}}{\delta g^{\mu\nu}}$.
Constant curvature background ($\bar{R}_{\mu\nu}=\Lambda\bar{g}_{\mu\nu}$)
should satisfy source-free equations of motion with the results\begin{equation}
\bar{K}=\bar{R}_{\mu\nu}^{2}-\frac{1}{2}\bar{R}^{2}=-4\Lambda^{2},\qquad\bar{S}=8\bar{R}^{\mu\nu}\bar{R}_{\mu\alpha}\bar{R}_{\nu}^{\alpha}-6\bar{R}\bar{R}_{\mu\nu}^{2}+\bar{R}^{3}=0,\end{equation}
 Then, the equations are satisfied if $\Lambda=\Lambda_{0}.$ Now,
let us linearize (\ref{eq:EOM_of_OR3_of_Rmn_act}) around its vacuum
(defining $T_{\mu\nu}\left(h\right)\equiv\delta\left(\frac{\tau_{\mu\nu}}{2}\right)$)
by use of the formulae in Appendix C and \begin{equation}
\delta K=-2\Lambda R_{L},\qquad\delta S=0.\end{equation}
 The linearized equations of motion after using the source-free equation
of motion for constant curvature background becomes\begin{equation}
T_{\mu\nu}\left(h\right)=\left(\frac{1}{\kappa}+\frac{4\alpha\Lambda_{0}}{3\kappa}\right)\mathcal{G}_{\mu\nu}^{L}-\frac{\alpha}{2\kappa}\left(\bar{\Box}\mathcal{G}_{\mu\nu}^{L}-\frac{2\Lambda_{0}}{3}\bar{g}_{\mu\nu}R_{L}\right),\end{equation}
 which exactly matches the equations that result from the matter coupled
version of the action (\ref{eq:OR2_of_Rmn_action}) as promised. This
shows explicitly that $O\left[\left(\alpha R\right)^{3}\right]$ terms
cancel each other. This cancellation will work for any arbitrary order
beyond this, as we will show with a different method below.

\subsubsection{Another method for unitarity analysis}

Hindawi \emph{et al }\cite{Hindawi} gave another method of analyzing
a generic higher derivative gravity model by reducing it to the equivalent
quadratic curvature theory in the sense that it has the same free
Lagrangian as the original higher derivative theory. Here, we will
review their approach and apply it to our example (\ref{eq:Rmn_action}).
Before we describe their method, we should note that unlike our method
which led to the compact formula (\ref{eq:Oh2_action}) that works
in all cases, the method of Hindawi \emph{et al} works only when one
deals with not matrices but scalar objects or one has a finite number
of curvature terms. Keeping this caveat in mind, which will be better
understood below, when Hindawi \emph{et al} method works, it provides
a fast algorithm in getting the equivalent quadratic action. 

To understand the essence of the Hindawi \emph{et al }method let us
consider the following simplified problem. Suppose we have a function
$f\left(x\left(t\right)\right)$, and we would like to find the $\epsilon^{2}$
order of $f\left(x\left(t_{0}+\epsilon\right)\right)$. But, instead
of doing this, we can find a function $g\left(x\left(t\right)\right)=a_{0}+a_{1}x\left(t\right)+a_{2}x^{2}\left(t\right)$
whose second order expansion around $t_{0}$ yields the same second
order expansion of $f\left(x\left(t\right)\right)$ around the same
point. After some straightforward analysis, one can show that $g\left(x\left(t\right)\right)$
can be obtained by expanding $f\left(x\left(t\right)\right)$ around
$x_{0}=x\left(t_{0}\right)$ up to and including $O\left[\left(x\left(t\right)-x_{0}\right)^{2}\right]$,
since $O\left[\left(x\left(t\right)-x_{0}\right)^{2+n}\right]$ gives
$\epsilon^{2+n}$ corrections with $n\ge1$. Hence, one can read the
coefficients for the correct $g\left(x\left(t\right)\right)$ to be
\begin{equation}
a_{0}=f\left(x_{0}\right)-\left[\frac{df}{dx}\right]_{x_{0}}x_{0}+\frac{1}{2}\left[\frac{d^{2}f}{dx^{2}}\right]_{x_{0}}x_{0}^{2},\qquad a_{1}=\left[\frac{df}{dx}\right]_{x_{0}}-x_{0}\left[\frac{d^{2}f}{dx^{2}}\right]_{x_{0}},\qquad a_{2}=\frac{1}{2}\left[\frac{d^{2}f}{dx^{2}}\right]_{x_{0}}.\end{equation}
 Note that if one just wants the {}``equations of motion,'' then
one carries out the above procedure at $O\left(\epsilon\right)$.
In this example, $f$ represents the Lagrangian, $x$ any curvature
tensor or scalar, and $\epsilon$ represents the metric perturbation
$h_{\mu\nu}\left(x\right)$. Similarly, $t_{0}$, $x_{0}$ are used
in analogy with the background metric $\bar{g}_{\mu\nu}$, etc.

\paragraph{Cubic theory}

Now, let us turn to our example (\ref{eq:Rmn_action}) and to specifically
its third order expansion in curvature given in (\ref{eq:OR3_of_Rmn_action}).
In order not to introduce the metric or its inverse during the expansion
around $\left(\bar{R},\bar{R}_{\nu}^{\mu}\right)$, let us take the
Lagrangian density of (\ref{eq:OR3_of_Rmn_action}) to be a function
of $R$ and $R_{\nu}^{\mu}$ as \begin{equation}
f\left(R,R_{\nu}^{\mu}\right)\equiv\frac{\alpha}{2}\left(R-2\Lambda_{0}\right)-\frac{\alpha^{2}}{4}\left(R_{\nu}^{\mu}R_{\mu}^{\nu}-\frac{1}{2}R^{2}\right)+\frac{\alpha^{3}}{48}\left(8R_{\rho}^{\mu}R_{\nu}^{\rho}R_{\mu}^{\nu}-6R_{\nu}^{\mu}R_{\mu}^{\nu}R+R^{3}\right).\label{eq:fRRmn_at_OR3}\end{equation}
 Expanding $f\left(R,R_{\nu}^{\mu}\right)$ around $\left(\bar{R},\bar{R}_{\nu}^{\mu}\right)$
with the assumption of small fluctuations about the background yields\begin{align}
g_{\text{quad-equal}}\left(R,R_{\nu}^{\mu}\right)= & f\left(\bar{R},\bar{R}_{\nu}^{\mu}\right)+\left[\frac{\partial f}{\partial R}\right]_{\left(\bar{R},\bar{R}_{\nu}^{\mu}\right)}\left(R-\bar{R}\right)+\left[\frac{\partial f}{\partial R_{\beta}^{\alpha}}\right]_{\left(\bar{R},\bar{R}_{\nu}^{\mu}\right)}\left(R_{\beta}^{\alpha}-\bar{R}_{\beta}^{\alpha}\right)\nonumber \\
 & +\frac{1}{2}\left[\frac{\partial^{2}f}{\partial R^{2}}\right]_{\left(\bar{R},\bar{R}_{\nu}^{\mu}\right)}\left(R-\bar{R}\right)^{2}+\left[\frac{\partial f}{\partial R\partial R_{\beta}^{\alpha}}\right]_{\left(\bar{R},\bar{R}_{\nu}^{\mu}\right)}\left(R-\bar{R}\right)\left(R_{\beta}^{\alpha}-\bar{R}_{\beta}^{\alpha}\right)\label{eq:Second_order_around_Rbars}\\
 & +\frac{1}{2}\left[\frac{\partial^{2}f}{\partial R_{\sigma}^{\rho}\partial R_{\beta}^{\alpha}}\right]_{\left(\bar{R},\bar{R}_{\nu}^{\mu}\right)}\left(R_{\beta}^{\alpha}-\bar{R}_{\beta}^{\alpha}\right)\left(R_{\sigma}^{\rho}-\bar{R}_{\sigma}^{\rho}\right).\nonumber \end{align}
 Computing the relevant derivatives one ends up with \begin{align}
g_{\text{quad-equal}}\left(R,R_{\nu}^{\mu}\right)= & f\left(\bar{R},\bar{R}_{\nu}^{\mu}\right)+\left[\frac{\alpha}{2}+\frac{\alpha^{2}}{4}\bar{R}+\frac{\alpha^{3}}{16}\left(\bar{R}^{2}-2\bar{R}_{\nu}^{\mu}\bar{R}_{\mu}^{\nu}\right)\right]\left(R-\bar{R}\right)\nonumber \\
 & +\left[-\frac{\alpha^{2}}{2}\bar{R}_{\alpha}^{\beta}+\frac{\alpha^{3}}{4}\left(2\bar{R}_{\nu}^{\beta}\bar{R}_{\alpha}^{\nu}-\bar{R}_{\alpha}^{\beta}\bar{R}\right)\right]\left(R_{\beta}^{\alpha}-\bar{R}_{\beta}^{\alpha}\right)\nonumber \\
 & +\frac{1}{2}\left(\frac{\alpha^{2}}{4}+\frac{\alpha^{3}}{8}\bar{R}\right)\left(R-\bar{R}\right)^{2}\\
 & +\left(-\frac{\alpha^{3}}{4}\bar{R}_{\alpha}^{\beta}\right)\left(R-\bar{R}\right)\left(R_{\beta}^{\alpha}-\bar{R}_{\beta}^{\alpha}\right)\nonumber \\
 & +\frac{1}{2}\left[-\frac{\alpha^{2}}{2}\delta_{\rho}^{\beta}\delta_{\alpha}^{\sigma}+\frac{\alpha^{3}}{4}\left(2\delta_{\rho}^{\beta}\bar{R}_{\alpha}^{\sigma}+2\bar{R}_{\rho}^{\beta}\delta_{\alpha}^{\sigma}-\delta_{\rho}^{\beta}\delta_{\alpha}^{\sigma}\bar{R}\right)\right]\left(R_{\beta}^{\alpha}-\bar{R}_{\beta}^{\alpha}\right)\left(R_{\sigma}^{\rho}-\bar{R}_{\sigma}^{\rho}\right).\nonumber \end{align}
 For constant curvature backgrounds, the corresponding quadratic action
becomes \begin{equation}
I=\frac{2}{\kappa\alpha}\int d^{4}x\sqrt{-g}\,\left[\frac{\alpha}{2}\left(R-2\Lambda_{0}\right)+\frac{\alpha^{2}}{8}R^{2}-\frac{\alpha^{2}}{4}R_{\mu\nu}^{2}\right],\end{equation}
 which once again shows that the cubic term in (\ref{eq:fRRmn_at_OR3})
does not contribute to the free theory. We should stress that if one
takes arbitrary coefficients instead of the ones we have which are
$(8,-6,1)$ at the cubic order (\ref{eq:fRRmn_at_OR3}), then one
would get a different quadratic action that does not follow from the
$A_{\mu\nu}$ (in this case it is just $\alpha R_{\mu\nu}$) expansion
of (\ref{eq:Rmn_action}). 

Now, let us also obtain the source-free nonlinear equations of motion
for a constant curvature background by finding the equivalent action
at $O\left(R\right)$. Similar steps lead to \begin{align}
g_{\text{lin-equal}}\left(R,R_{\nu}^{\mu}\right)= & f\left(\bar{R},\bar{R}_{\nu}^{\mu}\right)+\left[\frac{\partial f}{\partial R}\right]_{\left(\bar{R},\bar{R}_{\nu}^{\mu}\right)}\left(R-\bar{R}\right)+\left[\frac{\partial f}{\partial R_{\beta}^{\alpha}}\right]_{\left(\bar{R},\bar{R}_{\nu}^{\mu}\right)}\left(R_{\beta}^{\alpha}-\bar{R}_{\beta}^{\alpha}\right),\label{eq:First_order_around_in_Rbars}\end{align}
 and to the action \begin{equation}
I=\int d^{4}x\sqrt{-g}\,\left[\frac{\left(1+\alpha\Lambda\right)}{\kappa}\left(R-2\frac{\Lambda_{0}+\alpha\Lambda^{2}}{1+\alpha\Lambda}\right)+O\left(R^{2}\right)\right],\end{equation}
 where we used $\bar{R}_{\mu\nu}=\Lambda\bar{g}_{\mu\nu}$. Then,
identifying $\Lambda=\frac{\Lambda_{0}+\alpha\Lambda^{2}}{1+\alpha\Lambda}$,
one obtains $\Lambda=\Lambda_{0}.$

\paragraph{Full nonlinear action}

We mentioned above that Hindawi \emph{et al} method does not work
when one deals directly with matrices. Let us show this with \begin{equation}
f\left(R_{\mu\nu}\right)=\sqrt{\det\left(g_{\mu\nu}+\alpha A_{\mu\nu}\right)},\end{equation}
 and try to find $df$ which is needed for this analysis\emph{.} Defining
$M_{\mu\nu}\equiv g_{\mu\nu}+\alpha A_{\mu\nu}$, one has \emph{\begin{align}
df & =d\left(\sqrt{\det M}\right)=\frac{\sqrt{\det M}}{2}\text{Tr}\left[M^{-1}dM\right],\end{align}
 }where $M^{-1}dM$ is an ordinary matrix multiplication. Here, the
basic problem is to find $M^{-1}$ which cannot be done in exact form
for a general $A_{\mu\nu}$ and even when $A_{\mu\nu}=R_{\mu\nu}$.
But, one can always expand the determinant in terms of traces and
apply the Hindawi \emph{et al} method. Even though this is the case,
for a complicated $A_{\mu\nu}$ the determinant will yield many terms
in generic dimensions and as we show below even for four dimensions.
Let us consider the action (\ref{eq:Rmn_action}) and use the exact
formula\begin{equation}
\det M=\frac{1}{24}\left\{ \left(\text{Tr}M\right)^{4}-6\text{Tr}\left(M^{2}\right)\left(\text{Tr}M\right)^{2}+3\left[\text{Tr}\left(M^{2}\right)\right]^{2}+8\text{Tr}\left(M^{3}\right)\text{Tr}M-6\text{Tr}\left(M^{4}\right)\right\} .\end{equation}
 for $M=\delta_{\nu}^{\mu}+\alpha R_{\nu}^{\mu}$ , one gets \begin{align}
\det\left(\delta_{\nu}^{\mu}+\alpha R_{\nu}^{\mu}\right)= & 1+\alpha R+\frac{\alpha^{2}}{2}R^{2}+\frac{\alpha^{3}}{6}R^{3}+\frac{\alpha^{4}}{24}R^{4}-\frac{\alpha^{2}}{2}R_{\nu}^{\mu}R_{\mu}^{\nu}-\frac{\alpha^{3}}{2}RR_{\nu}^{\mu}R_{\mu}^{\nu}\nonumber \\
 & +\frac{\alpha^{3}}{3}R_{\rho}^{\mu}R_{\nu}^{\rho}R_{\mu}^{\nu}-\frac{\alpha^{4}}{4}R^{2}R_{\nu}^{\mu}R_{\mu}^{\nu}+\frac{\alpha^{4}}{3}RR_{\rho}^{\mu}R_{\nu}^{\rho}R_{\mu}^{\nu}+\frac{\alpha^{4}}{8}R_{\nu}^{\mu}R_{\mu}^{\nu}R_{\sigma}^{\rho}R_{\rho}^{\sigma}-\frac{\alpha^{4}}{4}R_{\rho}^{\mu}R_{\sigma}^{\rho}R_{\nu}^{\sigma}R_{\mu}^{\nu}.\label{eq:Exact_det_exp}\end{align}
 Defining $f\left(R,R_{\nu}^{\mu}\right)\equiv\sqrt{\det\left(\delta_{\nu}^{\mu}+\alpha R_{\nu}^{\mu}\right)}$
and with the help of (\ref{eq:Second_order_around_Rbars}) and the
formulae in Appendix D, we have the corresponding quadratic action
as\begin{equation}
I=\frac{2}{\kappa\alpha}\int d^{4}x\sqrt{-g}\,\left[\frac{\alpha}{2}\left(R-2\Lambda_{0}\right)+\frac{\alpha^{2}}{8}R^{2}-\frac{\alpha^{2}}{4}R_{\mu\nu}^{2}\right].\end{equation}
 Once again we have proven that the $O\left(R^{2+n}\right)$ with
$n\ge1$ terms do not contribute to the free theory for the exact
BI action (\ref{eq:Rmn_action}) around its constant curvature vacuum.
Note that with the help of an equivalent action at the linear level
as we have done before, \begin{align}
I & =\int d^{4}x\sqrt{-g}\,\left\{ \frac{1}{\kappa}\left(1+\alpha\Lambda\right)\left[R-2\left(\frac{\Lambda_{0}+\alpha\Lambda^{2}}{1+\alpha\Lambda}\right)\right]\right\} ,\end{align}
 setting $\Lambda=\frac{\Lambda_{0}+\alpha\Lambda^{2}}{1+\alpha\Lambda}$,
one has $\Lambda=\Lambda_{0}$. We should stress that to get this
result with the conventional method of finding the field equations
and looking for a solution of the form $\bar{R}_{\mu\nu}=\Lambda\bar{g}_{\mu\nu}$
is highly cumbersome for an action which is given as the square root
of $\det\left(\delta_{\nu}^{\mu}+\alpha R_{\nu}^{\mu}\right)$ (\ref{eq:Exact_det_exp}).

\subsection{Unitarity of the theory proposed by Deser and Gibbons}

While constructing the BI-type gravity actions, among various criteria,
one of the easiest to realize is the unitarity of the model around
flat space. This means when small curvature expansion is carried out
at the quadratic order in four dimensions, one should get the unique
theory $\frac{1}{\kappa}\left(R-2\Lambda_{0}\right)+\alpha R^{2}+\gamma\left(R^{2}-4R_{\mu\nu}^{2}+R_{\mu\nu\rho\sigma}^{2}\right)$
which is free of ghosts. Deser and Gibbons \cite{gibbonsDeser} suggested
that at the quadratic order, one should get, dropping the $\alpha R^{2}$
term, only the Einstein plus the Gauss-Bonnet combination (the $\gamma$
term). We will study such actions in a separate work, but here let
us consider an example (the one suggested by Deser and Gibbons) of
these models in which one does not have quadratic terms when expanded
around small curvature: \begin{equation}
I=\frac{2}{\kappa\alpha}\int d^{4}x\,\left[\sqrt{-\det\left[g_{\mu\nu}+\alpha R_{\mu\nu}+\frac{\alpha^{2}}{2}\left(R_{\mu\rho}R_{\nu}^{\rho}-\frac{1}{2}RR_{\mu\nu}\right)\right]}-\left(\alpha\Lambda_{0}+1\right)\sqrt{-\det g}\right].\label{eq:DeserGibbons_action}\end{equation}
 It is easy to see that the lowest order correction to the Einstein-Hilbert
theory goes like $O\left(R^{3}\right)$, which means around flat space
the graviton propagator is the same as that of Einstein-Hilbert theory.
(Note that for flat space to be the vacuum, one also sets $\Lambda_{0}=0$.)
But, around its constant curvature vacuum unitarity of this model
has not been checked before, since it is a highly nontrivial computation
without the tools we have developed above. To carry out the analysis,
we can find the $O\left(A_{\mu\nu}^{2}\right)$ action which has the
same $O\left(h^{2}\right)$ action as (\ref{eq:DeserGibbons_action}).
Here, $A_{\mu\nu}=\alpha R_{\mu\nu}+\frac{\alpha^{2}}{2}\left(R_{\mu\rho}R_{\nu}^{\rho}-\frac{1}{2}RR_{\mu\nu}\right)$
and let us stress again that $O\left(A_{\mu\nu}^{2}\right)$ action
is not equivalent to $O\left[\left(\alpha R\right)^{4}\right]$ action.
If one naively does the latter expansion, one will simply get an inconclusive
result since one would have neglected the $O\left[\left(\alpha R\right)^{4+n}\right]$
corrections. But, an expansion in $A_{\mu\nu}$ takes care of all
the relevant terms and cancellations. Therefore, using (\ref{eq:Small_M_expansion})
we have \begin{align}
I_{O\left(A^{2}\right)}=\frac{2}{\kappa\alpha}\int d^{4}x\,\sqrt{-g} & \left[\frac{\alpha}{2}\left(R-2\Lambda_{0}\right)+\frac{\alpha^{3}}{4}\left(RR_{\nu}^{\mu}R_{\mu}^{\nu}-R_{\rho}^{\mu}R_{\nu}^{\rho}R_{\mu}^{\nu}-\frac{1}{4}R^{3}\right)\right]\nonumber \\
 & \left.+\frac{\alpha^{4}}{32}\left(R_{\nu}^{\mu}R_{\mu}^{\nu}R_{\sigma}^{\rho}R_{\rho}^{\sigma}-\frac{3}{2}R^{2}R_{\nu}^{\mu}R_{\mu}^{\nu}+\frac{1}{4}R^{4}-2R_{\rho}^{\mu}R_{\nu}^{\rho}R_{\sigma}^{\nu}R_{\mu}^{\sigma}+2RR_{\rho}^{\mu}R_{\nu}^{\rho}R_{\mu}^{\nu}\right)\right]\end{align}
 Now, let us just concentrate on the higher curvature terms and define
\begin{align}
f\left(R,R_{\nu}^{\mu}\right)\equiv & \frac{\alpha^{3}}{4}\left(RR_{\nu}^{\mu}R_{\mu}^{\nu}-R_{\rho}^{\mu}R_{\nu}^{\rho}R_{\mu}^{\nu}-\frac{1}{4}R^{3}\right)\nonumber \\
 & +\frac{\alpha^{4}}{32}\left(R_{\nu}^{\mu}R_{\mu}^{\nu}R_{\sigma}^{\rho}R_{\rho}^{\sigma}-\frac{3}{2}R^{2}R_{\nu}^{\mu}R_{\mu}^{\nu}+\frac{1}{4}R^{4}-2R_{\rho}^{\mu}R_{\nu}^{\rho}R_{\sigma}^{\nu}R_{\mu}^{\sigma}+2RR_{\rho}^{\mu}R_{\nu}^{\rho}R_{\mu}^{\nu}\right),\label{eq:f_def_of_Deser_Gibbons_act}\end{align}
 The first thing we should find is the correct $\Lambda$ which can
be found by using the first order expansion (\ref{eq:First_order_around_in_Rbars})
of $f\left(R,R_{\nu}^{\mu}\right)$ around $\left(\bar{R},\bar{R}_{\nu}^{\mu}\right)=\left(4\Lambda,\Lambda\delta_{\nu}^{\mu}\right)$.
This procedure leads to the equivalent linear action\begin{equation}
I=\frac{2}{\kappa\alpha}\int d^{4}x\sqrt{-g}\,\left[\left(\frac{\alpha}{2}-\frac{3\alpha^{3}\Lambda^{2}}{4}+\frac{\alpha^{4}\Lambda^{3}}{4}\right)\left[R-2\frac{\left(\frac{\alpha\Lambda_{0}}{2}-\alpha^{3}\Lambda^{3}+\frac{3\alpha^{4}\Lambda^{4}}{8}\right)}{\left(\frac{\alpha}{2}-\frac{3\alpha^{3}\Lambda^{2}}{4}+\frac{\alpha^{4}\Lambda^{3}}{4}\right)}\right]+O\left(R^{2}\right)\right],\end{equation}
 from which one can get the equation that determines $\Lambda$ \begin{equation}
-\Lambda_{0}+\Lambda+\frac{\alpha^{2}\Lambda^{3}}{2}-\frac{\alpha^{3}\Lambda^{4}}{4}=0,\label{eq:eqn_of_mot_for_DeserGibbons_act}\end{equation}
 which has real roots, but they are not particularly illuminating
to display here. (One thing we can note is that even for $\Lambda_{0}=0$,
there are two real roots one of which is nonzero with a value $\Lambda\approx2.59/\alpha$.)
Now, we can employ the Hindawi \emph{et al} method to get the equivalent
quadratic action using (\ref{eq:Second_order_around_Rbars}) and the
relevant results of Appendix D: \begin{align}
I=\frac{2}{\kappa\alpha}\int d^{4}x\sqrt{-g} & \left[\left(-\alpha\Lambda_{0}-\alpha^{3}\Lambda^{3}+\frac{3\alpha^{4}\Lambda^{4}}{4}\right)+\left(\frac{\alpha}{2}+\frac{3\alpha^{3}\Lambda^{2}}{4}-\frac{\alpha^{4}\Lambda^{3}}{2}\right)R\right.\nonumber \\
 & \left.-\frac{\alpha^{3}\Lambda}{4}\left(1-\frac{\alpha\Lambda}{2}\right)R^{2}+\frac{\alpha^{3}\Lambda}{4}\left(1-\frac{\alpha\Lambda}{2}\right)R_{\mu\nu}^{2}\right].\end{align}
 For generic $\alpha$, this theory is plagued with a massive ghost
\cite{Stelle,Gullu1}. Thus, the action proposed by Deser and Gibbons
\cite{gibbonsDeser} does not yield a unitary spin-2 theory around
its constant curvature background for any choice of the curvature
except the flat space. But, setting $\alpha=\frac{2}{\Lambda}$ one
can get rid of the {}``bad'' $R_{\mu\nu}^{2}$ term, and hope to
obtain a unitary theory. However, this turns out to be not true, since
in this case setting $\Lambda=\Lambda_{0}$ which follows from (\ref{eq:eqn_of_mot_for_DeserGibbons_act}),
one ends up with\begin{equation}
I=\int d^{4}x\sqrt{-g}\left[-\frac{1}{\kappa}\left(R-2\Lambda_{0}\right)\right],\label{eq:a_equals_2/L_limit}\end{equation}
 which has the opposite sign of the Einstein-Hilbert action. That
means as long as one has $\kappa>0$ (which we must have for the unitarity
in flat space), the small fluctuations will have negative kinetic
energy and even for the tuned value of $\alpha$, (\ref{eq:DeserGibbons_action})
defines a nonunitary theory. We should note in passing that this result
does not necessarily imply negative energy for the \emph{exact} nonvacuum
solutions such as black holes of (\ref{eq:DeserGibbons_action}).
We have not yet found the black hole solutions of this action, but
we can give an example in which small fluctuations around the vacuum
have negative energy yet the exact solutions have positive energy.
This example is the Einstein-Gauss-Bonnet theory whose exact spherically
symmetric solution was given in \cite{Boulware} and whose energy
was computed in \cite{DeserTekin}. As discussed in the latter work,
this energy is positive, even though the linearized action of the
Einstein-Gauss Bonnet theory around its constant curvature vacuum
is opposite to that of Einstein's theory {[}just like (\ref{eq:a_equals_2/L_limit}){]}.
This is because the spherically symmetric Schwarzschild-de-Sitter
solution goes (say in five dimensions) as $-g_{00}=g^{rr}\sim1+\frac{m}{r^{2}}+\Lambda r^{2}$
unlike the usual Schwarzschild solution which goes like $-g_{00}=g^{rr}\sim1-\frac{m}{r^{2}}$,
the two minus signs take care of each other.

\section{Conclusion}

We have developed techniques of analyzing the unitarity of Born-Infeld
gravity actions around their constant curvature vacua. The special
determinantal form of the action gave rise to remarkable simplifications
that allow one to write a compact expression for the free, that is
$O\left(h^{2}\right)$, theory. To summarize our result, let us note
the following: One needs to find the $O\left(h^{2}\right)$ action
of \begin{equation}
\mathcal{L}=\frac{2}{\kappa\alpha}\left[\sqrt{-\det\left(\delta_{\nu}^{\mu}+\alpha R_{\nu}^{\mu}+\beta\left(\text{\text{Riem}, Ric, R, ...}\right)_{\nu}^{\mu}\right)}-\left(\alpha\Lambda_{0}+1\right)\right],\label{eq:BI_Lagrangian}\end{equation}
 to study its tree-level unitarity. In this work what we have done
is to give a method to determine the parameters $K$, $\Lambda$,
$a$, $b$, $c$ in the following Lagrangian whose $O\left(h^{2}\right)$
expansion equals that of (\ref{eq:BI_Lagrangian}) \begin{align*}
\mathcal{L}_{\text{equivalent}}= & \frac{1}{K\left(\kappa,\alpha,\beta,\Lambda_{0},\dots\right)}\left[R-2\Lambda\left(\kappa,\alpha,\beta,\Lambda_{0},\dots\right)\right]\\
 & +a\left(\kappa,\alpha,\beta,\Lambda_{0},\dots\right)R^{2}+b\left(\kappa,\alpha,\beta,\Lambda_{0},\dots\right)R_{\mu\nu}^{2}+c\left(\kappa,\alpha,\beta,\Lambda_{0},\dots\right)R_{\mu\nu\rho\sigma}^{2},\end{align*}
 to \emph{all orders} in the curvature expansion. Once this equivalent
quadratic Lagrangian is obtained, unitarity analysis proceeds with
the standard methods as discussed in \cite{Gullu1}. We have also
presented two examples one of which was proposed as a unitary theory
in flat space \cite{gibbonsDeser}, but turned out to be nonunitary
in curved space according to our computation above. The other simpler
example was considered to show the details of our method.

Let us give a recipe of how one should check the tree-level unitarity
of a given Born-Infeld gravity in generic dimension $D$ around its
constant curvature vacuum. First to find the effective cosmological
constant $\Lambda$, one has to expand the action up to $O\left(h\right)$
around the constant curvature vacuum using (\ref{eq:Oh_action}),
or one should find the equivalent linear, that is $O\left(R\right)$,
action and read the cosmological constant from it. Then, one should
find the $O\left(h^{2}\right)$ action using (\ref{eq:Oh2_action})
or alternatively one should construct the equivalent quadratic action
{[}$O\left(R^{2}\right)${]}. The method we have presented (\ref{eq:Oh2_action})
works just as good in odd and even dimensions. But, the second method,
as discussed in detail in the text, which proceeds by construction
of an equivalent $O\left(R\right)$ and $O\left(R^{2}\right)$ actions
should be done with great care depending on the number of dimensions
and on the complexity of $A_{\mu\nu}$. The original BI action cannot
simply be expanded in small curvature to get these equivalent actions
via (\ref{eq:Small_M_expansion}). What always works, in principle,
is that the determinant can be expanded exactly in terms of traces
within the square root, then one can do the expansions (\ref{eq:First_order_around_in_Rbars})
and (\ref{eq:Second_order_around_Rbars}), and use the Hindawi \emph{et
al }technique. But, the exact expansion of the determinant in terms
of traces can generate quite a large number of terms especially for
$D\ge4$. {[}For example, in (\ref{eq:DeserGibbons_action}) doing
such an exact expansion is not advised to the reader.{]} Therefore,
to get the equivalent action one should proceed as follows in generic
even dimensions $D=2n+2$: one has to expand the BI action up to $O\left(A^{n+1}\right)$
with $n\ge1$ using (\ref{eq:Small_M_expansion}), if the resultant
action is not already quadratic in the curvature, then using (\ref{eq:Second_order_around_Rbars}),
the equivalent quadratic action should be constructed. For generic
odd dimensions, the best way is to use (\ref{eq:Oh2_action}), but
for $D=3$ and for not so complicated $A_{\mu\nu}$, exact trace expansion
can also be employed. In this work we have laid out the details of
checking unitarity of BI gravities, in a separate work we will provide
examples of unitary models around flat and constant curvature backgrounds
\cite{Gullu0}.

\section{\label{ackno} Acknowledgments}

This work is supported by the T{Ü}B\.{I}TAK Grant No. 110T339, and
METU Grant No. BAP-07-02-2010-00-02. Some of the calculations in this
paper were either done or checked with the help of the computer package
Cadabra \cite{Cadabra1,Cadabra2}.

\section*{Appendix A: A Two-Dimensional Example}

In order to understand why in even dimensions finite number of terms
in the $A_{\mu\nu}$ expansion of the BI-type actions contribute to
$O\left(h\right)$ and $O\left(h^{2}\right)$ expansions, let us study
a simple two-dimensional determinantal function \begin{equation}
f\left(\tau,\gamma\right)=\sqrt{\det\left[\left(\begin{array}{cc}
1 & 0\\
0 & 1\end{array}\right)+\gamma\left(\begin{array}{cc}
a\left(\tau\right) & b\left(\tau\right)\\
c\left(\tau\right) & d\left(\tau\right)\end{array}\right)\right]},\label{eq:f}\end{equation}
 where $\tau$ and $\gamma$ are two independent variables. The $\tau$,
$\gamma$ expansions of $f\left(\tau,\gamma\right)$ represent the
metric perturbation expansion and the $A_{\mu\nu}$ expansion, respectively,
for the BI-type actions. What we will show in this Appendix is that
$f\left(\tau,\gamma\right)$ and the function $g\left(\tau,\gamma\right)$
defined as \begin{equation}
g\left(\tau,\gamma\right)\equiv1+\frac{1}{2}\gamma\left[a\left(\tau\right)+d\left(\tau\right)\right],\label{eq:g}\end{equation}
 have the same the $O\left(\tau\right)$ expansion around $\tau=0$
only if \begin{equation}
\left[\left(\begin{array}{cc}
a\left(\tau\right) & b\left(\tau\right)\\
c\left(\tau\right) & d\left(\tau\right)\end{array}\right)\right]_{\tau=0}=\left(\begin{array}{cc}
a_{0} & 0\\
0 & a_{0}\end{array}\right),\label{eq:t0_matrix}\end{equation}
 which is the analog of the maximally symmetric constant curvature
background in the BI-type gravity. Here, the important point about
$g\left(\tau,\gamma\right)$ is that it is just the $O\left(\gamma\right)$
expansion of $f\left(\tau,\gamma\right)$ obtained by using (\ref{eq:Small_M_expansion}),
but note that we exactly define $g\left(\tau,\gamma\right)$ in this
way and \emph{do not assume that $\gamma$ is small}. Thus, staying
at first order in $\tau$ expansion requires just the first order
in $\gamma$, while one naively expects that first order in $\tau$
expansion should involve each order in $\gamma$. Let us understand
this in more detail by considering a generic function $\phi\left(\tau,\gamma\right)$
and expand it in $\tau$ as a Taylor series around $\tau=0$\begin{equation}
\phi\left(\tau,\gamma\right)=\phi\left(\tau=0,\gamma\right)+\left(\frac{\partial\phi}{\partial\tau}\right)_{\tau=0}\tau+O\left(\tau^{2}\right)+\dots,\end{equation}
 where $\left(\frac{\partial\phi}{\partial\tau}\right)_{\tau=0}$
is a function of $\gamma$ only. One can write the power series expansion
of $\left(\frac{\partial\phi}{\partial\tau}\right)_{\tau=0}$ in $\gamma$
by assuming $\phi\left(\tau,\gamma\right)=\sum_{i=0}^{\infty}\psi_{i}\left(\tau\right)\gamma^{i}$
and expanding each $\psi_{i}\left(\tau\right)$ to the first order
in $\tau$. Then, one has \begin{align}
\phi\left(\tau,\gamma\right) & =\sum_{i=0}^{\infty}\psi_{i}\left(\tau=0\right)\gamma^{i}+\left[\sum_{i=0}^{\infty}\left(\frac{\partial\psi_{i}}{\partial\tau}\right)_{\tau=0}\gamma^{i}\right]\tau+\dots\Rightarrow\left(\frac{\partial\phi}{\partial\tau}\right)_{\tau=0}=\sum_{i=0}^{\infty}\left(\frac{\partial\psi_{i}}{\partial\tau}\right)_{\tau=0}\gamma^{i}.\end{align}
 For the determinantal function $f\left(\tau,\gamma\right)$, the
terms $\left(\frac{\partial\psi_{i}}{\partial\tau}\right)_{\tau=0},\, i\ge2$
are all zero. Let us observe this for the $i=2$ term explicitly.
First, one can have the $O\left(\gamma^{2}\right)$ expansion of $f\left(\tau,\gamma\right)$
by using (\ref{eq:Small_M_expansion}) as\begin{align}
\sqrt{\det\left[\left(\begin{array}{cc}
1 & 0\\
0 & 1\end{array}\right)+\gamma\left(\begin{array}{cc}
a\left(\tau\right) & b\left(\tau\right)\\
c\left(\tau\right) & d\left(\tau\right)\end{array}\right)\right]} & =1+\frac{1}{2}\gamma\left[a\left(\tau\right)+d\left(\tau\right)\right]+\frac{1}{8}\gamma^{2}\left[a\left(\tau\right)+d\left(\tau\right)\right]^{2}\nonumber \\
 & \phantom{=}-\frac{1}{4}\gamma^{2}\left[a^{2}\left(\tau\right)+2b\left(\tau\right)c\left(\tau\right)+d^{2}\left(\tau\right)\right]+O\left(\gamma^{3}\right).\end{align}
 Assuming \begin{equation}
\left(\begin{array}{cc}
a\left(\tau\right) & b\left(\tau\right)\\
c\left(\tau\right) & d\left(\tau\right)\end{array}\right)=\left(\begin{array}{cc}
a_{0} & 0\\
0 & a_{0}\end{array}\right)+\left(\begin{array}{cc}
a_{1} & b_{1}\\
c_{1} & d_{1}\end{array}\right)\tau+O\left(\tau^{2}\right),\label{eq:t_exp_of_matrix}\end{equation}
 one has\begin{align}
f\left(\tau,\gamma\right) & =1+\frac{1}{2}\gamma\left[\left(a_{0}+\tau a_{1}\right)+\left(a_{0}+\tau d_{1}\right)\right]+\frac{1}{8}\gamma^{2}\left[\left(a_{0}+\tau a_{1}\right)+\left(a_{0}+\tau d_{1}\right)\right]^{2}\nonumber \\
 & \phantom{=}-\frac{1}{4}\gamma^{2}\left[\left(a_{0}+\tau a_{1}\right)^{2}+2\left(\tau b_{1}\right)\left(\tau c_{1}\right)+\left(a_{0}+\tau d_{1}\right)^{2}\right]+O\left(\gamma^{3}\right)\nonumber \\
 & =\left(1+\gamma a_{0}\right)+\frac{1}{2}\gamma\tau\left(a_{1}+d_{1}\right)+\frac{1}{8}\gamma^{2}\left[4a_{0}^{2}+4\tau a_{0}\left(a_{1}+d_{1}\right)+O\left(\tau^{2}\right)\right]\\
 & \phantom{=}-\frac{1}{4}\gamma^{2}\left[2a_{0}^{2}+2\tau a_{0}\left(a_{1}+d_{1}\right)+O\left(\tau^{2}\right)\right]+O\left(\gamma^{3}\right)\nonumber \\
 & =\left(1+\gamma a_{0}\right)+\frac{1}{2}\gamma\tau\left(a_{1}+d_{1}\right)+O\left(\tau^{2}\right)+O\left(\gamma^{3}\right).\nonumber \end{align}
 Thus, $O\left(\tau\right)$ contributions coming from the two $O\left(\gamma^{2}\right)$
terms cancel each other because of the specific coefficients in (\ref{eq:Small_M_expansion})
and the assumption (\ref{eq:t0_matrix}). Now, let us verify our proposal
by explicitly calculating $O\left(\tau\right)$ expansions of $f\left(\tau,\gamma\right)$
and $g\left(\tau,\gamma\right)$. By using (\ref{eq:t_exp_of_matrix})
in $f\left(\tau,\gamma\right)$, one obtains \begin{multline}
\sqrt{\det\left[\left(\begin{array}{cc}
1 & 0\\
0 & 1\end{array}\right)+\gamma\left(\begin{array}{cc}
a_{0}+\tau a_{1} & \tau b_{1}\\
\tau c_{1} & a_{0}+\tau d_{1}\end{array}\right)+O\left(\tau^{2}\right)\right]}\\
=\left(1+\gamma a_{0}\right)\sqrt{\det\left[\left(\begin{array}{cc}
1 & 0\\
0 & 1\end{array}\right)+\frac{\gamma\tau}{\left(1+\gamma a_{0}\right)}\left(\begin{array}{cc}
a_{1} & b_{1}\\
c_{1} & d_{1}\end{array}\right)+O\left(\tau^{2}\right)\right]},\end{multline}
 and it is possible to make the $O\left(\tau\right)$ expansion by
using (\ref{eq:Small_M_expansion}); \begin{align}
f\left(\tau,\gamma\right) & =\left(1+\gamma a_{0}\right)\left[1+\frac{1}{2}\frac{\gamma\tau}{\left(1+\gamma a_{0}\right)}\left(a_{1}+d_{1}\right)+O\left(\tau^{2}\right)\right]=\left(1+\gamma a_{0}\right)+\frac{1}{2}\gamma\tau\left(a_{1}+d_{1}\right)+O\left(\tau^{2}\right).\end{align}
 Therefore, the $O\left(\gamma^{2+n}\right),\, n\ge1$ terms in $\gamma$
expansion of $f\left(\tau,\gamma\right)$ do not contribute to the
$O\left(\tau\right)$ terms, only if (\ref{eq:t0_matrix}) holds.
On the other hand, the $O\left(\tau\right)$ expansion of $g\left(\tau,\gamma\right)$
can be simply found as \begin{equation}
g\left(\tau,\gamma\right)=1+\frac{1}{2}\gamma\left[2a_{0}+\tau\left(a_{1}+d_{1}\right)\right]=\left(1+\gamma a_{0}\right)+\frac{1}{2}\gamma\tau\left(a_{1}+d_{1}\right).\end{equation}
 As a result, if one wants to consider $O\left(\tau\right)$ behavior
of $f\left(\tau,\gamma\right)$, then one can equally work with just
$g\left(\tau,\gamma\right)$ which is simply equal to the $O\left(\gamma\right)$
expansion of $f\left(\tau,\gamma\right)$.

\section*{Appendix B: Analyzing Einstein-Hilbert Action and Quadratic Curvature
Gravity with Second Order Perturbations}

In this Appendix, second order expansions of the curvature tensors
are used in the well-known cases of the Einstein-Hilbert theory, and
the quadratic actions including the Einstein-Gauss-Bonnet theory.
This will help us construct the following $O\left(h^{2}\right)$ actions
that frequently appear in the computations \[
\int d^{4}x\,\sqrt{-\bar{g}}R_{\left(2\right)},\qquad\int d^{4}x\,\sqrt{-\bar{g}}\bar{g}^{\mu\nu}R_{\mu\nu}^{\left(2\right)},\qquad\int d^{4}x\,\sqrt{-\bar{g}}R_{L}^{\mu\nu}R_{\mu\nu}^{L},\]
 \begin{equation}
\int d^{4}x\,\sqrt{-\bar{g}}\left(R_{\mu\rho\sigma\lambda}^{2}\right)^{\left(2\right)},\qquad\int d^{4}x\,\sqrt{-\bar{g}}\bar{g}^{\sigma\nu}\bar{g}^{\lambda\gamma}\left(R_{\phantom{\mu}\rho\sigma\lambda}^{\mu}\right)^{\left(1\right)}\left(R_{\phantom{\rho}\mu\gamma\nu}^{\rho}\right)^{\left(1\right)},\end{equation}
 in terms of the building blocks appearing in Eq. (25) of \cite{DeserTekin}.

\subsection*{Analysis of the Einstein-Hilbert action}

First, let us find the second order in metric perturbation for Einstein-Hilbert
action:\begin{equation}
I=\frac{1}{\kappa}\int d^{4}x\,\sqrt{-g}\left(R-2\Lambda_{0}\right),\end{equation}
 and expanding up to third order in $h_{\mu\nu}$ yields\begin{align}
I=\frac{1}{\kappa}\int d^{4}x\,\sqrt{-\bar{g}} & \left[1+\frac{\tau}{2}h+\frac{1}{8}\tau^{2}\left(h^{2}-2h_{\mu\nu}^{2}\right)+O\left(\tau^{3}\right)\right]\left[\left(\bar{R}-2\Lambda_{0}\right)+\tau R_{L}+\tau^{2}R_{\left(2\right)}+O\left(\tau^{3}\right)\right]\nonumber \\
=\frac{1}{\kappa}\int d^{4}x\,\sqrt{-\bar{g}} & \Biggl\{\left(\bar{R}-2\Lambda_{0}\right)+\tau\left[\frac{1}{2}h\left(\bar{R}-2\Lambda_{0}\right)+R_{L}\right]\nonumber \\
 & +\tau^{2}\left[\frac{1}{8}\left(\bar{R}-2\Lambda_{0}\right)\left(h^{2}-2h_{\mu\nu}^{2}\right)+\frac{1}{2}hR_{L}+R_{\left(2\right)}\right]+O\left(\tau^{3}\right)\Biggr\}.\end{align}

One can find the nonlinear equation of motion for constant curvature
background by investigating the first order term in $\tau$ of the
above action as\begin{equation}
I_{O\left(h\right)}=\frac{1}{\kappa}\int d^{4}x\,\sqrt{-\bar{g}}\left[\frac{1}{2}h\left(\bar{R}-2\Lambda_{0}\right)+R_{L}\right],\end{equation}
 after putting the explicit form of $R_{L}$ and dropping out a boundary
term one can get \begin{equation}
I_{O\left(h\right)}=\frac{1}{\kappa}\int d^{4}x\,\sqrt{-\bar{g}}h\left(\Lambda-\Lambda_{0}\right),\end{equation}
 from which it follows that $\left(\Lambda-\Lambda_{0}\right)\bar{g}_{\mu\nu}=0$
upon taking variation with respect to $h_{\mu\nu}$. 

One can read the second order action as \begin{equation}
I_{O\left(h^{2}\right)}=\frac{1}{\kappa}\int d^{4}x\,\sqrt{-\bar{g}}\left\{ h^{\mu\nu}\left[\frac{1}{2}\left(\Lambda-\frac{1}{2}\Lambda_{0}\right)\left(\bar{g}_{\mu\nu}h-2h_{\mu\nu}\right)+\frac{1}{2}\bar{g}_{\mu\nu}R_{L}\right]+R_{\left(2\right)}\right\} \label{eq:EH_Oh2}\end{equation}
 where $R_{\left(2\right)}$ can be read from (\ref{eq:Second_order_R})
as\begin{equation}
R_{\left(2\right)}=\bar{R}^{\rho\lambda}h_{\alpha\rho}h_{\lambda}^{\alpha}-h^{\mu\nu}R_{\mu\nu}^{L}-\bar{g}^{\nu\sigma}h_{\beta}^{\mu}\left(R_{\phantom{\mu}\nu\mu\sigma}^{\beta}\right)_{L}-\bar{g}^{\nu\sigma}\bar{g}^{\mu\alpha}\bar{g}_{\beta\gamma}\left[\left(\Gamma_{\mu\alpha}^{\gamma}\right)_{L}\left(\Gamma_{\sigma\nu}^{\beta}\right)_{L}-\left(\Gamma_{\sigma\alpha}^{\gamma}\right)_{L}\left(\Gamma_{\mu\nu}^{\beta}\right)_{L}\right].\end{equation}
 Let us concentrate on $\int d^{4}x\,\sqrt{-\bar{g}}R_{\left(2\right)}$
part of the action and work out the integration by parts;\begin{align}
\int d^{4}x\,\sqrt{-\bar{g}}R_{\left(2\right)}=\int d^{4}x\,\sqrt{-\bar{g}} & \Biggl\{\Lambda h_{\mu\nu}^{2}-h^{\mu\nu}R_{\mu\nu}^{L}-\bar{g}^{\nu\sigma}h_{\beta}^{\mu}\left(R_{\phantom{\mu}\nu\mu\sigma}^{\beta}\right)_{L}\nonumber \\
 & -\bar{g}^{\nu\sigma}\bar{g}^{\mu\alpha}\bar{g}_{\beta\gamma}\left[\left(\Gamma_{\mu\alpha}^{\gamma}\right)_{L}\left(\Gamma_{\sigma\nu}^{\beta}\right)_{L}-\left(\Gamma_{\sigma\alpha}^{\gamma}\right)_{L}\left(\Gamma_{\mu\nu}^{\beta}\right)_{L}\right]\Biggr\}.\end{align}
 One can find $\bar{g}^{\nu\sigma}h_{\beta}^{\mu}\left(R_{\phantom{\mu}\nu\mu\sigma}^{\beta}\right)_{L}$
as\begin{equation}
\bar{g}^{\nu\sigma}h_{\beta}^{\mu}\left(R_{\phantom{\mu}\nu\mu\sigma}^{\beta}\right)_{L}=h^{\mu\nu}\left(R_{\mu\nu}^{L}-\frac{4\Lambda}{3}h_{\mu\nu}+\frac{\Lambda}{3}\bar{g}_{\mu\nu}h\right).\label{eq:ghRiemann}\end{equation}
 By using the definition of the linearized Christoffel connection
in (\ref{eq:Linear_Christoffel}) and doing integration by parts,
the last two terms in $\int d^{4}x\,\sqrt{-\bar{g}}R_{\left(2\right)}$
can be found as \begin{align}
\int d^{4}x\,\sqrt{-\bar{g}}\bar{g}^{\nu\sigma}\bar{g}^{\mu\alpha}\bar{g}_{\beta\gamma}\left(\Gamma_{\mu\alpha}^{\gamma}\right)_{L}\left(\Gamma_{\sigma\nu}^{\beta}\right)_{L} & =\int d^{4}x\,\sqrt{-\bar{g}}\left[-\frac{1}{2}h^{\mu\nu}\left(\bar{\nabla}^{\sigma}\bar{\nabla}_{\mu}h_{\nu\sigma}+\bar{\nabla}^{\sigma}\bar{\nabla}_{\nu}h_{\mu\sigma}-\frac{3}{2}\bar{\nabla}_{\mu}\bar{\nabla}_{\nu}h\right)\right.\nonumber \\
 & \phantom{=\int d^{4}x\,\sqrt{-\bar{g}}}\left.+h^{\mu\nu}\left(\frac{4\Lambda}{3}h_{\mu\nu}-\frac{\Lambda}{12}\bar{g}_{\mu\nu}h\right)+\frac{1}{4}h^{\mu\nu}\bar{g}_{\mu\nu}R_{L}\right],\end{align}
 \begin{equation}
\int d^{4}x\,\sqrt{-\bar{g}}\bar{g}^{\nu\sigma}\bar{g}^{\mu\alpha}\bar{g}_{\beta\gamma}\left(\Gamma_{\sigma\alpha}^{\gamma}\right)_{L}\left(\Gamma_{\mu\nu}^{\beta}\right)_{L}=\int d^{4}x\,\sqrt{-\bar{g}}\left[-\frac{1}{4}h^{\mu\nu}\left(3\bar{\Box}h_{\mu\nu}-\bar{\nabla}^{\sigma}\bar{\nabla}_{\mu}h_{\sigma\nu}-\bar{\nabla}^{\sigma}\bar{\nabla}_{\nu}h_{\mu\sigma}\right)\right].\end{equation}
 Finally, $\int d^{4}x\,\sqrt{-\bar{g}}R_{\left(2\right)}$ becomes
\begin{equation}
\int d^{4}x\,\sqrt{-\bar{g}}R_{\left(2\right)}=\int d^{4}x\,\sqrt{-\bar{g}}h^{\mu\nu}\left(-\frac{1}{2}R_{\mu\nu}^{L}-\frac{1}{4}\bar{g}_{\mu\nu}R_{L}+\Lambda h_{\mu\nu}-\frac{\Lambda}{4}\bar{g}_{\mu\nu}h\right),\label{eq:Integral_R2}\end{equation}
 and putting this result in (\ref{eq:EH_Oh2}) yields\begin{equation}
I_{O\left(h^{2}\right)}=-\frac{1}{2\kappa}\int d^{4}x\,\sqrt{-\bar{g}}h^{\mu\nu}\left[\mathcal{G}_{\mu\nu}^{L}+\frac{1}{2}\left(\Lambda_{0}-\Lambda\right)\left(\bar{g}_{\mu\nu}h-2h_{\mu\nu}\right)\right],\end{equation}
 and since $\Lambda=\Lambda_{0}$ is found from equations of motion
for constant curvature background;\begin{equation}
I_{O\left(h^{2}\right)}=-\frac{1}{2\kappa}\int d^{4}x\,\sqrt{-\bar{g}}h^{\mu\nu}\mathcal{G}_{\mu\nu}^{L}.\end{equation}

\subsection*{Analysis of the quadratic action}

Now, let us consider the quadratic actions in the form\begin{equation}
I=\int d^{4}x\,\sqrt{-g}\left[\frac{1}{\kappa}\left(R-2\Lambda_{0}\right)+\alpha R^{2}+\beta R_{\mu\nu}^{2}\right],\end{equation}
 and calculate the second order action in metric perturbations. Then,
up to third order, the expansion of the action is\begin{align}
I=\int d^{4}x\, & \sqrt{-\bar{g}}\left\{ \left[\frac{1}{\kappa}\left(\bar{R}-2\Lambda_{0}\right)+\alpha\bar{R}^{2}+\beta\bar{R}_{\mu\nu}^{2}\right]\right.\nonumber \\
 & \left.+\tau\left[\frac{1}{2}h\left(\frac{1}{\kappa}\left(\bar{R}-2\Lambda_{0}\right)+\alpha\bar{R}^{2}+\beta\bar{R}_{\mu\nu}^{2}\right)\right.\right.\nonumber \\
 & \left.\phantom{+\tau}\left.+\left(\frac{1}{\kappa}R_{L}+2\alpha\bar{R}R_{L}+\beta\bar{R}^{\mu\nu}R_{\mu\nu}^{L}+\beta\left(R^{\mu\nu}\right)_{\left(1\right)}\bar{R}_{\mu\nu}\right)\right]\right.\\
 & \left.+\tau^{2}\left[\frac{1}{8}\left(h^{2}-2h_{\mu\nu}^{2}\right)\left(\frac{1}{\kappa}\left(\bar{R}-2\Lambda_{0}\right)+\alpha\bar{R}^{2}+\beta\bar{R}_{\mu\nu}^{2}\right)\right.\right.\nonumber \\
 & \left.\phantom{+\tau^{2}}\left.+\frac{1}{2}h\left(\frac{1}{\kappa}R_{L}+2\alpha\bar{R}R_{L}+\beta\bar{R}^{\mu\nu}R_{\mu\nu}^{L}+\beta\left(R^{\mu\nu}\right)_{\left(1\right)}\bar{R}_{\mu\nu}\right)\right.\right.\nonumber \\
 & \left.\phantom{+\tau^{2}}\left.+\left(\frac{1}{\kappa}R_{\left(2\right)}+2\alpha\bar{R}R_{\left(2\right)}+\alpha R_{L}^{2}+\beta\bar{R}^{\mu\nu}R_{\mu\nu}^{\left(2\right)}+\beta\left(R^{\mu\nu}\right)_{\left(1\right)}R_{\mu\nu}^{L}+\beta\left(R^{\mu\nu}\right)_{\left(2\right)}\bar{R}_{\mu\nu}\right)\right]\right\} .\nonumber \end{align}
 Here, note that $\left(R^{\mu\nu}\right)_{\left(1\right)}$ and $\left(R^{\mu\nu}\right)_{\left(2\right)}$
are the first and the second order terms in the metric perturbation
expansion of $R^{\mu\nu}$. First of all, in order to find the nonlinear
equation of motion for constant curvature background, one needs to
study the $O\left(\tau\right)$ term in the above action. After using
the definitions of $R_{\mu\nu}^{L}$, $R_{L}$ and dropping out the
boundary terms one can get\begin{equation}
I_{O\left(h\right)}=\frac{1}{\kappa}\int d^{4}x\,\sqrt{-\bar{g}}h\left(\Lambda-\Lambda_{0}\right),\end{equation}
 which yields the equation of motion $\Lambda=\Lambda_{0}$. Then,
let us move to the second order term in metric perturbation. After
using the result given in (\ref{eq:Integral_R2}), one can obtain
\begin{align}
I_{O\left(h^{2}\right)}=-\frac{1}{2}\int d^{4}x\,\sqrt{-\bar{g}} & \left\{ \left(\frac{1}{\kappa}+8\alpha\Lambda+4\beta\Lambda\right)h^{\mu\nu}\mathcal{G}_{\mu\nu}^{L}\right.\nonumber \\
 & \left.-\frac{1}{2}h^{2}\left[\frac{1}{\kappa}\left(\Lambda-\Lambda_{0}\right)+2\beta\Lambda^{2}\right]+h_{\mu\nu}^{2}\left[\frac{1}{\kappa}\left(\Lambda-\Lambda_{0}\right)+6\beta\Lambda^{2}\right]\right.\\
 & \left.+2\alpha h^{\mu\nu}\left(\bar{g}_{\mu\nu}\bar{\Box}-\bar{\nabla}_{\mu}\bar{\nabla}_{\nu}+\Lambda\bar{g}_{\mu\nu}\right)R_{L}\right.\nonumber \\
 & \left.-2\beta\left(\Lambda\bar{g}^{\mu\nu}R_{\mu\nu}^{\left(2\right)}+R_{L}^{\mu\nu}R_{\mu\nu}^{L}+R_{\left(2\right)}^{\mu\nu}\Lambda\bar{g}_{\mu\nu}\right)\right\} .\nonumber \end{align}
 Here, let us first handle $\int d^{4}x\,\sqrt{-\bar{g}}R_{L}^{\mu\nu}R_{\mu\nu}^{L}$.
Using the definition of $R_{\mu\nu}^{L}$ and using the linearized
Bianchi identity (and also its covariant derivative) which is \begin{equation}
\bar{\nabla}^{\mu}\mathcal{G}_{\mu\nu}^{L}=0,\qquad\mathcal{G}_{\mu\nu}^{L}\equiv R_{\mu\nu}^{L}-\frac{1}{2}\bar{g}_{\mu\nu}R_{L}-\Lambda h_{\mu\nu},\end{equation}
 one can find the following result after use of integration by parts
\begin{align}
\int d^{4}x\,\sqrt{-\bar{g}}R_{L}^{\mu\nu}R_{\mu\nu}^{L}=-\frac{1}{2}\int d^{4}x\,\sqrt{-\bar{g}}h^{\mu\nu} & \left[\left(\bar{g}_{\mu\nu}\bar{\Box}-\bar{\nabla}_{\mu}\bar{\nabla}_{\nu}+\Lambda\bar{g}_{\mu\nu}\right)R_{L}+\left(\bar{\Box}\mathcal{G}_{\mu\nu}^{L}-\frac{2\Lambda}{3}\bar{g}_{\mu\nu}R_{L}\right)\right.\nonumber \\
 & \left.-\frac{14\Lambda}{3}R_{\mu\nu}^{L}+\frac{\Lambda}{3}\bar{g}_{\mu\nu}R_{L}+\frac{8\Lambda^{2}}{3}h_{\mu\nu}\right].\end{align}
 Second, $\left(R^{\mu\nu}\right)_{\left(2\right)}$ is related to
$R_{\mu\nu}^{\left(2\right)}$ in the following way:\begin{equation}
\bar{g}_{\mu\nu}\left(R^{\mu\nu}\right)_{\left(2\right)}=\bar{g}_{\mu\nu}\left(g^{\mu\alpha}g^{\nu\beta}R_{\alpha\beta}\right)^{\left(2\right)}=\bar{g}^{\mu\nu}R_{\mu\nu}^{\left(2\right)}-2h^{\mu\nu}R_{\mu\nu}^{L}+3\Lambda h_{\mu\nu}^{2}.\end{equation}
 The $\bar{g}^{\mu\nu}R_{\mu\nu}^{\left(2\right)}$ term can be given
in terms of $R_{\left(2\right)}$ with\begin{equation}
R_{\left(2\right)}=\left(g^{\mu\nu}R_{\mu\nu}\right)_{\left(2\right)}=\bar{g}^{\mu\nu}R_{\mu\nu}^{\left(2\right)}-h^{\mu\nu}R_{\mu\nu}^{L}+\Lambda h_{\mu\nu}^{2}.\end{equation}
 Then, $\int d^{4}x\,\sqrt{-\bar{g}}\bar{g}^{\mu\nu}R_{\mu\nu}^{\left(2\right)}$
becomes\begin{equation}
\int d^{4}x\,\sqrt{-\bar{g}}\bar{g}^{\mu\nu}R_{\mu\nu}^{\left(2\right)}=h^{\mu\nu}\left(\frac{1}{2}R_{\mu\nu}^{L}-\frac{1}{4}\bar{g}_{\mu\nu}R_{L}-\frac{\Lambda}{4}\bar{g}_{\mu\nu}h\right),\end{equation}
 with the help of (\ref{eq:Integral_R2}). By use of these results
and the equation of motion for constant curvature background which
is $\Lambda=\Lambda_{0}$ in $I_{O\left(h^{2}\right)}$, one can get\begin{align}
I_{O\left(h^{2}\right)}=-\frac{1}{2}\int d^{4}x\,\sqrt{-\bar{g}}h^{\mu\nu} & \left[\left(\frac{1}{\kappa}+8\alpha\Lambda+\frac{4}{3}\beta\Lambda\right)\mathcal{G}_{\mu\nu}^{L}\right.\nonumber \\
 & \left.+\left(2\alpha+\beta\right)\left(\bar{g}_{\mu\nu}\bar{\Box}-\bar{\nabla}_{\mu}\bar{\nabla}_{\nu}+\Lambda\bar{g}_{\mu\nu}\right)R_{L}+\beta\left(\bar{\Box}\mathcal{G}_{\mu\nu}^{L}-\frac{2\Lambda}{3}\bar{g}_{\mu\nu}R_{L}\right)\right],\end{align}
 which is same as Eq. (25) of \cite{DeserTekin}.

Finally, let us analyze the Einstein-Gauss-Bonnet theory,\begin{equation}
I=\int d^{4}x\,\sqrt{-g}\left[\frac{1}{\kappa}\left(R-2\Lambda_{0}\right)+\gamma\left(R_{\mu\rho\sigma\lambda}^{2}-4R_{\mu\nu}^{2}+R^{2}\right)\right],\end{equation}
 just to check the consistency of our construction. Here, the only
remaining part that we have not analyzed is the $R_{\mu\rho\sigma\lambda}^{2}$
term. First, let us use the previous result in order to obtain the
second order action in metric perturbations for the terms other than
$R_{\mu\rho\sigma\lambda}^{2}$:\begin{align}
I_{O\left(h^{2}\right)}=-\frac{1}{2}\int d^{4}x\,\sqrt{-\bar{g}}h^{\mu\nu} & \biggl[\left(\frac{1}{\kappa}+\frac{8}{3}\gamma\Lambda\right)\mathcal{G}_{\mu\nu}^{L}-2\gamma\left(\bar{g}_{\mu\nu}\bar{\Box}-\bar{\nabla}_{\mu}\bar{\nabla}_{\nu}+\Lambda\bar{g}_{\mu\nu}\right)R_{L}\nonumber \\
 & -4\gamma\left(\bar{\Box}\mathcal{G}_{\mu\nu}^{L}-\frac{2\Lambda}{3}\bar{g}_{\mu\nu}R_{L}\right)\biggr].\label{eq:R2_Ricci2_part}\end{align}
 Then, up to third order, expansion of the last term becomes\begin{align}
I=\gamma\int d^{4}x\,\sqrt{-\bar{g}} & \Biggl\{\bar{R}_{\mu\rho\sigma\lambda}^{2}+\tau\left[\left(R_{\mu\rho\sigma\lambda}^{2}\right)^{\left(1\right)}+\frac{1}{2}h\bar{R}_{\mu\rho\sigma\lambda}^{2}\right].\nonumber \\
 & +\tau^{2}\left[\left(R_{\mu\rho\sigma\lambda}^{2}\right)^{\left(2\right)}+\frac{1}{2}h\left(R_{\mu\rho\sigma\lambda}^{2}\right)^{\left(1\right)}+\frac{1}{8}\bar{R}_{\mu\rho\sigma\lambda}^{2}\left(h^{2}-2h_{\mu\nu}^{2}\right)\right]\Biggr\}.\end{align}
 First of all, it should be shown that first order part is a boundary
term such that it should not give a contribution to equation of motion
for constant curvature background:\begin{equation}
I_{O\left(h\right)}=\int d^{4}x\,\sqrt{-\bar{g}}\left[\left(R_{\mu\rho\sigma\lambda}^{2}\right)^{\left(1\right)}+\frac{1}{2}h\bar{R}_{\mu\rho\sigma\lambda}^{2}\right],\end{equation}
 where \begin{equation}
\bar{R}_{\mu\rho\sigma\lambda}^{2}=\frac{8\Lambda^{2}}{3},\qquad\left(R_{\mu\rho\sigma\lambda}^{2}\right)^{\left(1\right)}=\frac{4\Lambda}{3}R_{L}.\end{equation}
 Then,\begin{align}
I_{O\left(h\right)} & =\int d^{4}x\,\sqrt{-\bar{g}}\left[\left(\bar{\nabla}^{\mu}\bar{\nabla}^{\nu}h_{\mu\nu}-\bar{\Box}h\right)\right],\end{align}
 and since the remaining part is a boundary term, no contribution
comes to the constant curvature background equation of motion from
the square of the Riemann tensor. Then, moving to the part that is
second order in metric perturbation\begin{equation}
I_{O\left(h^{2}\right)}=\gamma\int d^{4}x\,\sqrt{-\bar{g}}\left[\left(R_{\mu\rho\sigma\lambda}^{2}\right)^{\left(2\right)}+\frac{2\Lambda}{3}hR_{L}+\frac{\Lambda^{2}}{3}\left(h^{2}-2h_{\mu\nu}^{2}\right)\right],\end{equation}
 where\begin{align}
\left(R_{\mu\rho\sigma\lambda}^{2}\right)^{\left(2\right)}= & \left(R_{\phantom{\mu}\rho\sigma\lambda}^{\mu}R_{\phantom{\rho}\mu\gamma\nu}^{\rho}g^{\sigma\nu}g^{\lambda\gamma}\right)^{\left(2\right)}\nonumber \\
= & 2\bar{R}_{\phantom{\rho}\mu}^{\rho\phantom{\mu}\lambda\sigma}\left(R_{\phantom{\mu}\rho\sigma\lambda}^{\mu}\right)^{\left(2\right)}+2\bar{R}_{\phantom{\mu}\rho\sigma\lambda}^{\mu}\bar{R}_{\phantom{\rho}\mu\phantom{\lambda}\nu}^{\rho\phantom{\mu}\lambda}g_{\left(2\right)}^{\sigma\nu}+\bar{g}^{\sigma\nu}\bar{g}^{\lambda\gamma}\left(R_{\phantom{\mu}\rho\sigma\lambda}^{\mu}\right)^{\left(1\right)}\left(R_{\phantom{\rho}\mu\gamma\nu}^{\rho}\right)^{\left(1\right)}\nonumber \\
 & +2\left[\bar{R}_{\phantom{\rho}\mu\phantom{\lambda}\nu}^{\rho\phantom{\mu}\lambda}\left(R_{\phantom{\mu}\rho\sigma\lambda}^{\mu}\right)^{\left(1\right)}+\bar{R}_{\phantom{\mu}\rho\sigma}^{\mu\phantom{\rho\sigma}\gamma}\left(R_{\phantom{\rho}\mu\gamma\nu}^{\rho}\right)^{\left(1\right)}\right]g_{\left(1\right)}^{\sigma\nu}+\bar{R}_{\phantom{\mu}\rho\sigma\lambda}^{\mu}\bar{R}_{\phantom{\rho}\mu\gamma\nu}^{\rho}g_{\left(1\right)}^{\sigma\nu}g_{\left(1\right)}^{\lambda\gamma},\end{align}
 Using $\bar{R}_{\mu\nu\rho\sigma}=\frac{\Lambda}{3}\left(\bar{g}_{\mu\rho}\bar{g}_{\nu\sigma}-\bar{g}_{\mu\sigma}\bar{g}_{\nu\rho}\right)$
and $R_{\left(2\right)}=\bar{g}^{\rho\sigma}R_{\rho\sigma}^{\left(2\right)}+g_{\left(1\right)}^{\rho\sigma}R_{\rho\sigma}^{\left(1\right)}+\bar{R}_{\rho\sigma}g_{\left(2\right)}^{\rho\sigma}$;\begin{equation}
\left(R_{\mu\rho\sigma\lambda}^{2}\right)^{\left(2\right)}=\frac{4\Lambda}{3}R_{\left(2\right)}+\bar{g}^{\sigma\nu}\bar{g}^{\lambda\gamma}\left(R_{\phantom{\mu}\rho\sigma\lambda}^{\mu}\right)^{\left(1\right)}\left(R_{\phantom{\rho}\mu\gamma\nu}^{\rho}\right)^{\left(1\right)}-\frac{4\Lambda}{3}\bar{g}^{\rho\lambda}\left(R_{\phantom{\mu}\rho\sigma\lambda}^{\mu}\right)^{\left(1\right)}h_{\mu}^{\sigma}+\frac{2\Lambda^{2}}{9}\left(h^{2}-h_{\mu\nu}^{2}\right),\end{equation}
 and using (\ref{eq:ghRiemann})\begin{align}
\left(R_{\mu\rho\sigma\lambda}^{2}\right)^{\left(2\right)}= & \frac{4\Lambda}{3}R_{\left(2\right)}-\bar{g}^{\sigma\nu}\bar{g}^{\lambda\gamma}\left(R_{\phantom{\mu}\rho\lambda\sigma}^{\mu}\right)^{\left(1\right)}\left(R_{\phantom{\rho}\mu\gamma\nu}^{\rho}\right)^{\left(1\right)}\nonumber \\
 & -\frac{4\Lambda}{3}h^{\mu\nu}R_{\mu\nu}^{L}+\frac{14\Lambda^{2}}{9}h_{\mu\nu}^{2}-\frac{2\Lambda^{2}}{9}h^{2}.\end{align}
 Now, let us consider $\int d^{4}x\,\sqrt{-\bar{g}}\left(R_{\mu\rho\sigma\lambda}^{2}\right)^{\left(2\right)}$.
The $\int d^{4}x\,\sqrt{-\bar{g}}\bar{g}^{\sigma\nu}\bar{g}^{\lambda\gamma}\left(R_{\phantom{\mu}\rho\sigma\lambda}^{\mu}\right)^{\left(1\right)}\left(R_{\phantom{\rho}\mu\gamma\nu}^{\rho}\right)^{\left(1\right)}$
term can be found as \begin{align}
\bar{g}^{\mu\nu}\bar{g}^{\rho\alpha}\left(R_{\phantom{\lambda}\sigma\rho\mu}^{\lambda}\right)^{\left(1\right)}\left(R_{\phantom{\sigma}\lambda\alpha\nu}^{\sigma}\right)^{\left(1\right)}= & h^{\mu\nu}\left[2\left(\bar{\Box}\mathcal{G}_{\mu\nu}^{L}-\frac{2\Lambda}{3}\bar{g}_{\mu\nu}R_{L}\right)+\left(\bar{g}_{\mu\nu}\bar{\Box}-\bar{\nabla}_{\mu}\bar{\nabla}_{\nu}+\Lambda\bar{g}_{\mu\nu}\right)R_{L}\right]\nonumber \\
 & -\frac{\Lambda}{9}h^{\mu\nu}\left(30R_{\mu\nu}^{L}-9\bar{g}_{\mu\nu}R_{L}-32\Lambda h_{\mu\nu}+2\Lambda\bar{g}_{\mu\nu}h\right),\end{align}
 after a somewhat lengthy calculation where the definition of the
linearized Riemann tensor is used and the terms are rearranged by
using integration by parts. Using this result with (\ref{eq:Integral_R2}),
one get\begin{align}
\int d^{4}x\,\sqrt{-\bar{g}}\left(R_{\mu\rho\sigma\lambda}^{2}\right)^{\left(2\right)}=\int d^{4}x\,\sqrt{-\bar{g}}h^{\mu\nu} & \Biggl\{-\left[2\left(\bar{\Box}\mathcal{G}_{\mu\nu}^{L}-\frac{2\Lambda}{3}\bar{g}_{\mu\nu}R_{L}\right)+\left(\bar{g}_{\mu\nu}\bar{\Box}-\bar{\nabla}_{\mu}\bar{\nabla}_{\nu}+\Lambda\bar{g}_{\mu\nu}\right)R_{L}\right]\nonumber \\
 & +\frac{\Lambda}{3}\left(8\mathcal{G}_{\mu\nu}^{L}-4R_{\mu\nu}^{L}+6\Lambda h_{\mu\nu}-\Lambda\bar{g}_{\mu\nu}h\right)\Biggr\},\end{align}
 and plugging it in the action:\begin{equation}
I_{O\left(h^{2}\right)}=-\frac{1}{2}\gamma\int d^{4}x\,\sqrt{-\bar{g}}h^{\mu\nu}\left[-\frac{8\Lambda}{3}\mathcal{G}_{\mu\nu}^{L}+2\left(\bar{g}_{\mu\nu}\bar{\Box}-\bar{\nabla}_{\mu}\bar{\nabla}_{\nu}+\Lambda\bar{g}_{\mu\nu}\right)R_{L}+4\left(\bar{\Box}\mathcal{G}_{\mu\nu}^{L}-\frac{2\Lambda}{3}\bar{g}_{\mu\nu}R_{L}\right)\right],\end{equation}
 and considering this result with the part of the action coming from
$\gamma R^{2}$ and $-4\gamma R_{\mu\nu}^{2}$ terms given in (\ref{eq:R2_Ricci2_part}),
one finds that all the $\gamma$ terms vanish, and the Gauss-Bonnet
term does not contribute to the equation of motion.

\section*{Appendix C: Linearization of the $O\left(R^{3}\right)$ Action}

The following formulae are needed for the linearization of the $O\left[\left(\alpha R\right)^{3}\right]$
equations. The quadratic parts below already appeared in \cite{DeserTekin},
we reproduce them here for the sake of completeness, the cubic parts
are new. \[
\delta\left(R_{\lambda\nu\alpha\mu}R^{\lambda\alpha}\right)=\frac{2\Lambda}{3}R_{\mu\nu}^{L}+\frac{\Lambda}{3}\bar{g}_{\mu\nu}R_{L}+\frac{\Lambda^{2}}{3}h_{\mu\nu},\]
 \[
\delta\left(\Box R_{\mu\nu}\right)=\bar{\Box}R_{\mu\nu}^{L}-\Lambda\bar{\Box}h_{\mu\nu},\qquad\delta\left(\nabla_{\mu}\nabla_{\nu}R\right)=\bar{\nabla}_{\mu}\bar{\nabla}_{\nu}R_{L},\qquad\delta\left(\Box R\right)=\bar{\Box}R_{L},\]
 \begin{equation}
\delta\left(R_{\mu}^{\rho}R_{\rho\alpha}R_{\nu}^{\alpha}\right)=3\Lambda^{2}R_{\mu\nu}^{L}-2\Lambda^{3}h_{\mu\nu},\qquad\delta\left(R_{\mu\nu}R_{\alpha\beta}^{2}\right)=4\Lambda^{2}R_{\mu\nu}^{L}+2\Lambda^{2}\bar{g}_{\mu\nu}R_{L},\end{equation}
 \[
\delta\left(R_{\mu\nu}R^{2}\right)=16\Lambda^{2}R_{\mu\nu}^{L}+8\Lambda^{2}\bar{g}_{\mu\nu}R_{L},\qquad\delta\left(RR_{\nu}^{\rho}R_{\mu\rho}\right)=8\Lambda^{2}R_{\mu\nu}^{L}+\Lambda^{2}\bar{g}_{\mu\nu}R_{L}-4\Lambda^{3}h_{\mu\nu},\]
 \[
\delta\left(\nabla_{\alpha}\nabla_{\mu}R_{\nu}^{\alpha}\right)=\frac{1}{2}\bar{\nabla}_{\mu}\bar{\nabla}_{\nu}R_{L}+\frac{4\Lambda}{3}R_{\mu\nu}^{L}-\frac{\Lambda}{3}\bar{g}_{\mu\nu}R_{L}-\frac{4\Lambda^{2}}{3}h_{\mu\nu},\]
 \[
\delta\left(\nabla_{\mu}\nabla_{\nu}R_{\alpha\beta}\right)=\bar{\nabla}_{\mu}\bar{\nabla}_{\nu}R_{\alpha\beta}-\Lambda\bar{\nabla}_{\mu}\bar{\nabla}_{\nu}h_{\alpha\beta}.\]
 Here, last two equations can related by using linearized Bianchi
identity: \begin{equation}
\bar{\nabla}^{\mu}\mathcal{G}_{\mu\nu}^{L}=0,\qquad\mathcal{G}_{\mu\nu}^{L}\equiv R_{\mu\nu}^{L}-\frac{1}{2}\bar{g}_{\mu\nu}R_{L}-\Lambda h_{\mu\nu}.\end{equation}

\section*{Appendix D: Coefficients for the $R-\bar{R}$ Expansion}

Coefficients in the expansion of the square root of (\ref{eq:Exact_det_exp})
are

\begin{align*}
\left[\frac{\partial f}{\partial R}\right]_{\left(\bar{R},\bar{R}_{\nu}^{\mu}\right)} & =\frac{\alpha}{2\left(1+\alpha\Lambda\right)^{2}}\left(1+4\alpha\Lambda+6\alpha^{2}\Lambda^{2}+4\alpha^{3}\Lambda^{3}\right),\end{align*}
 \begin{align*}
\left[\frac{\partial f}{\partial R_{\beta}^{\alpha}}\right]_{\left(\bar{R},\bar{R}_{\nu}^{\mu}\right)}= & -\frac{\alpha\delta_{\alpha}^{\beta}}{2\left(1+\alpha\Lambda\right)^{2}}\left(\alpha\Lambda+3\alpha^{2}\Lambda^{2}+3\alpha^{3}\Lambda^{3}\right),\end{align*}
 \begin{align}
\left[\frac{\partial^{2}f}{\partial R^{2}}\right]_{\left(\bar{R},\bar{R}_{\nu}^{\mu}\right)}= & \frac{\alpha^{2}}{2\left(1+\alpha\Lambda\right)^{2}}\left(1+4\alpha\Lambda+6\alpha^{2}\Lambda^{2}\right)-\frac{\alpha^{2}}{4\left(1+\alpha\Lambda\right)^{6}}\left(1+4\alpha\Lambda+6\alpha^{2}\Lambda^{2}+4\alpha^{3}\Lambda^{3}\right)^{2},\end{align}
 \begin{align*}
\left[\frac{\partial^{2}f}{\partial R_{\sigma}^{\rho}\partial R_{\beta}^{\alpha}}\right]_{\left(\bar{R},\bar{R}_{\nu}^{\mu}\right)}=-\frac{\alpha^{2}}{2\left(1+\alpha\Lambda\right)^{2}} & \left[\left(1+\alpha\Lambda\right)^{2}\delta_{\rho}^{\beta}\delta_{\alpha}^{\sigma}-\alpha^{2}\Lambda^{2}\delta_{\alpha}^{\beta}\delta_{\rho}^{\sigma}\right]\\
-\frac{\alpha^{2}\delta_{\alpha}^{\beta}\delta_{\rho}^{\sigma}}{4\left(1+\alpha\Lambda\right)^{6}} & \left(\alpha\Lambda+3\alpha^{2}\Lambda^{2}+3\alpha^{3}\Lambda^{3}\right)^{2},\end{align*}
 \begin{align*}
\left[\frac{\partial f}{\partial R\partial R_{\beta}^{\alpha}}\right]_{\left(\bar{R},\bar{R}_{\nu}^{\mu}\right)}= & -\frac{\alpha^{2}\delta_{\alpha}^{\beta}}{2\left(1+\alpha\Lambda\right)^{2}}\left(\alpha\Lambda+3\alpha^{2}\Lambda^{2}\right)\\
 & +\frac{\alpha^{2}\delta_{\alpha}^{\beta}}{4\left(1+\alpha\Lambda\right)^{6}}\left(1+4\alpha\Lambda+6\alpha^{2}\Lambda^{2}+4\alpha^{3}\Lambda^{3}\right)\left(\alpha\Lambda+3\alpha^{2}\Lambda^{2}+3\alpha^{3}\Lambda^{3}\right).\end{align*}

Coefficients in the expansion of the (\ref{eq:f_def_of_Deser_Gibbons_act})
are \[
f\left(\bar{R},\bar{R}_{\nu}^{\mu}\right)=-\alpha^{3}\Lambda^{3}\left(1-\frac{\alpha\Lambda}{4}\right),\qquad\left[\frac{\partial f}{\partial R}\right]_{\left(\bar{R},\bar{R}_{\nu}^{\mu}\right)}=-2\alpha^{3}\Lambda^{2}\left(1-\frac{3\alpha\Lambda}{8}\right),\]
 \begin{equation}
\left[\frac{\partial f}{\partial R_{\beta}^{\alpha}}\right]_{\left(\bar{R},\bar{R}_{\nu}^{\mu}\right)}=\frac{5\alpha^{3}\Lambda^{2}}{4}\delta_{\alpha}^{\beta}-\frac{\alpha^{4}\Lambda^{3}}{2}\delta_{\alpha}^{\beta},\qquad\left[\frac{\partial^{2}f}{\partial R^{2}}\right]_{\left(\bar{R},\bar{R}_{\nu}^{\mu}\right)}=-\frac{3\alpha^{3}}{2}\Lambda+\frac{9\alpha^{4}\Lambda^{2}}{8},\end{equation}
 \[
\left[\frac{\partial f}{\partial R\partial R_{\beta}^{\alpha}}\right]_{\left(\bar{R},\bar{R}_{\nu}^{\mu}\right)}=\frac{\alpha^{3}\Lambda}{2}\delta_{\alpha}^{\beta}-\frac{9\alpha^{4}\Lambda^{2}}{16}\delta_{\alpha}^{\beta},\qquad\left[\frac{\partial^{2}f}{\partial R_{\sigma}^{\rho}\partial R_{\beta}^{\alpha}}\right]_{\left(\bar{R},\bar{R}_{\nu}^{\mu}\right)}=\frac{\alpha^{3}\Lambda}{2}\delta_{\rho}^{\beta}\delta_{\alpha}^{\sigma}-\frac{\alpha^{4}\Lambda^{2}}{4}\delta_{\rho}^{\beta}\delta_{\alpha}^{\sigma}+\frac{\alpha^{4}\Lambda^{2}}{4}\delta_{\rho}^{\sigma}\delta_{\alpha}^{\beta}.\]

\end{document}